\documentclass[a4paper,10pt] {scrreport}
\usepackage[utf8]{inputenc}
\usepackage[greek,english]{babel}
\usepackage[T1]{fontenc}
\usepackage[export]{adjustbox} 
\usepackage{csquotes}
\usepackage[backend=bibtex]{biblatex}
\usepackage[left=2.9cm,right=2.9cm,top=3cm,bottom=4.5cm]{geometry}
\usepackage{amsmath}
\usepackage{graphicx}
\usepackage{float}
\usepackage{multirow}
\usepackage{longtable}
\usepackage{setspace}
\usepackage{comment}
\newcommand{\abs}[1]{\ensuremath{\left\vert#1\right\vert}}
\usepackage{authblk}
\usepackage{array}
\newcolumntype{C}[1]{>{\centering\arraybackslash}p{#1}}
\usepackage{enumerate}

\usepackage{xurl}
\usepackage{hyperref}
\hypersetup{breaklinks=true}

\title{Adaptive Shock Compensation in the Multi-layer Network of Global Food Production and Trade}

\author[1]{Sophia Baum\thanks{baum@csh.ac.at}}
\author[1,2]{Moritz Laber}
\author[3,4]{Martin Bruckner}
\author[1]{Liuhuaying Yang}
\author[1,5,6,7]{Stefan Thurner}
\author[1,5,7]{Peter Klimek}

\affil[1]{Complexity Science Hub Vienna, Josefstädterstraße 39, 1080 Vienna, Austria}
\affil[2]{Network Science Institute, Northeastern University, 360 Huntington Ave, Boston, MA 02115, USA}
\affil[3]{Institute of Environmental Engineering, ETH Zurich, Laura-Hezner-Weg 7, 8093 Zurich, Switzerland}
\affil[4]{Institute for Ecological Economics, WU Vienna, Welthandelsplatz 1, 1020 Vienna, Austria}
\affil[5]{Center for Medical Data Science CeDAS, Medical University of Vienna, Spitalgasse 23, 1080 Vienna, Austria}
\affil[6]{Santa Fe Institute, 1399 Hyde Park Road, Santa Fe, NM 87501, USA}
\affil[7]{Supply Chain Intelligence Institute Austria, Josefstädterstraße 39, 1080 Vienna, Austria}

\begin{document}

\maketitle 

\setcounter{page}{2} 

\begin{abstract}
\textbf{Abstract}\\

Global food production and trade networks are highly dynamic, especially in response to shortages when countries adjust their supply strategies. In this study, we examine adjustments across $123$ agri-food products from $192$ countries resulting in $23616$ individual scenarios of food shortage, and calibrate a multi-layer network model to understand the propagation of the shocks. We analyze shock mitigation actions, such as increasing imports, boosting production, or substituting food items. Our findings indicate that these lead to spillover effects potentially exacerbating food inequality: an Indian rice shock resulted in a $5.8$ \% increase in rice losses in countries with a low Human Development Index (HDI) and a $14.2$ \% decrease in those with a high HDI. Considering multiple interacting shocks leads to super-additive losses of up to $12$ \% of the total available food volume across the global food production network. This framework allows us to identify combinations of shocks that pose substantial systemic risks and reduce the resilience of the global food supply.\\
\\

\textbf{Keywords}\\

shock propagation, adaptive networks, multi-layer networks, temporal networks,

network dynamics, food security

\end{abstract}

\pagestyle{headings}

\chapter{Introduction}\label{Introduction}

The World Food Programme estimates that up to $309$ million people face acute food insecurity in $2024$ [\hyperlink{1}{1}]. In Somalia, $15.6$ out of $16.4$ million ($95.1$\%) people do not have access to sufficient amounts of food, in Afghanistan $77.0$\% and in Haiti $67.2$\% suffer the same fate [\hyperlink{2}{2}]. Other severely affected countries are located in the Sahel zone, Central Africa, the Americas, and West and Central Asia [\hyperlink{1}{1}], where especially conflicts have led to the most drastic episodes in the world food crisis [\hyperlink{1}{1}]. 

Geopolitical factors further threaten food security. For example, since $2022$, India, the world's largest rice exporter, introduced restrictions on grain exports, including a complete ban on non-basmati and broken white rice [\hyperlink{3}{3},\hyperlink{4}{4}]. Officially, the Indian government justifies the restrictions as necessary to protect domestic food security with a growing population and the looming effects of climate change [\hyperlink{3}{3}]. In response, over the past year, global and domestic rice prices rose substantially by double-digit percentage points, as India plays a substantial role in the global rice trade, accounting for about $32.6$\% and a trade volume of $9.7$ billion USD [\hyperlink{3}{3},\hyperlink{4}{4}].  

Russia exposed the world to another major threat to food security in February $2022$, when it invaded Ukraine. The conflict has had a severe political and economic impact on both countries. Ukraine, a major producer of maize, wheat, and sunflower oil, has suffered from the destruction of infrastructure and depletion of labour force \hyperlink{5}{5}. The disruption of grain shipments through the Black Sea has exacerbated the $2022$ global food price crisis \hyperlink{6}{6}. With the July $2022$ Black Sea Grain Initiative currently uncertain, exports remain vulnerable to Russian attacks \hyperlink{6}{6}.\\

Today, food production is embedded in a global trade network through which shortages propagate and can develop from a sudden local reduction in available food, i.e., a shock, to potential global food security crises, often through cascading [\hyperlink{7}{7},\hyperlink{8}{8},\hyperlink{9}{9}]. Shocks with such destructive potential have become more frequent in recent times \hyperlink{10}{10}. Quantitative methods for assessing food security are often based on modelling food prices [\hyperlink{11}{11},\hyperlink{12}{12},\hyperlink{13}{13},\hyperlink{14}{14}], typically with a focus on individual commodities or commodity groups and their respective trade networks [\hyperlink{8}{8},\hyperlink{15}{15},\hyperlink{16}{16},\hyperlink{17}{17},\hyperlink{18}{18}]. However, since many commodities, especially staples such as wheat, maize, soybeans, and rice, serve as inputs in production processes [\hyperlink{19}{19}], there is an increasing interest in shock propagation models that include indirect trade and production network effects [\hyperlink{8}{8},\hyperlink{15}{15}, \hyperlink{16}{16},\hyperlink{20}{20},\hyperlink{21}{21}] meaning that shortages of input products may spill over into several output products along the transformation chain [\hyperlink{20}{20}].
Traditionally, input-output models are used to capture the transformation of products into each other [\hyperlink{22}{22},\hyperlink{23}{23}]. While effective as demand-driven models for uncovering the upstream use of resources along the supply chain [\hyperlink{19}{19}], input-output models prove less suitable for assessing supply-driven shocks [\hyperlink{23}{23}].\\

When assessing the impact of a food supply shock on a country, the adaptive capacity of the global food trade and production network is typically neglected, with links in the food supply network treated as static [\hyperlink{8}{8},\hyperlink{9}{9},\hyperlink{20}{20}] even though the system changes over the years [\hyperlink{17}{17},\hyperlink{24}{24},\hyperlink{25}{25}].
During the 2008 food crisis, one quarter of $77$ states surveyed by the FAO took action to increase food supply \hyperlink{26}{26}. Maintaining fixed food trading and production strategies for shocked products in a crisis lead to exacerbated and possibly avoidable losses [\hyperlink{15}{15},\hyperlink{18}{18},\hyperlink{21}{21},\hyperlink{27}{27},\hyperlink{28}{28}]. However, adaptation efforts to mitigate one's own losses can cause spillover effects throughout the trading network due to intensified competition for fewer resources [\hyperlink{16}{16}]. In adjusting the trade and production network, a country has a wide variety of possibilities. Therefore, establishing country-specific adjustment strategies usually requires detailed knowledge of the country's production capabilities, geopolitical alliances, price formation, and financial possibilities  [\hyperlink{17}{17},\hyperlink{18}{18}].\\

In this study, we tackle the challenge of developing and validating country- and product-specific adaptation rules through a data-driven analysis. We achieve this by parameterizing a response strategy model based on extensive historical food input-output (IO) data [\hyperlink{19}{19} ]and numerous shocks and subsequent adaptations documented there. The modelled adaptation rules are integrated into a bipartite multilayer shock propagation model, which was originally formulated using static recursive trade and production functions [\hyperlink{20}{20}]. Our model explores $10^8$ potential shock transmission pathways. We extend this framework by incorporating adaptive, empirically grounded, and country-specific strategies for trade, production, allocation, and substitution that are triggered by substantial losses. This allows us to simulate global outcomes of shock scenarios for specific products in various countries and to evaluate and compare the effectiveness of different adaptation strategies across countries and products.\\

Our approach can be summarized as follows. First, we identify relevant shock events by examining annual variations in the availability of a specific product within each country. Second, we derive adjustment rules from the observed changes in trade and production parameters following these events. These rules redefine the trade and production functions set out by [\hyperlink{20}{20}], modifying existing trade and production relationships or establishing new ones. Finally, we apply these adjustment rules to hypothetical shocks in the recursive trade and production functions and analyze the results.\\

We utilize this framework to provide a comprehensive overview of historical food availability shocks, explore general patterns in country responses to these shocks, and identify the most effective adaptation rules. We investigate two specific hypothetical scenarios inspired by recent geopolitical events: a total embargo on Indian rice [\hyperlink{3}{3},\hyperlink{4}{4}] and a disruption of Ukrainian wheat exports due to the ongoing Russian invasion [\hyperlink{5}{5},\hyperlink{6}{6}]. For both of these scenarios, we consider a worst-case situation involving a complete loss of production and assess the resulting impact on global availability of the products as well as on secondary items. Additionally, we examine the combined effects of both shocks, given that rice and wheat can serve as substitutes for one another.\\
\\
\begin{figure}
  \centering
  \includegraphics[width=0.9\textwidth]{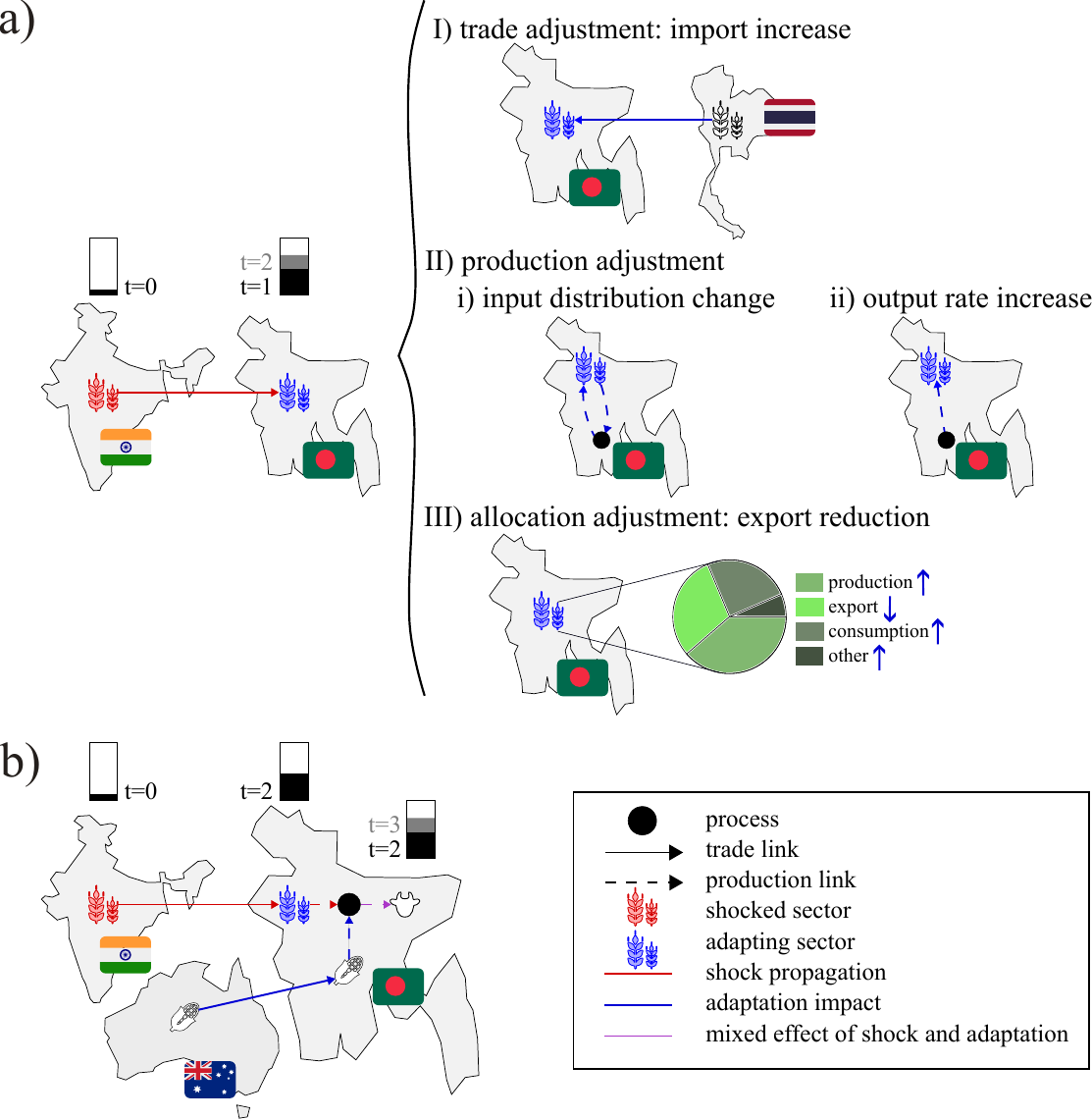}
  \label{model overview}
  \vspace{1cm}
  \caption{Schematic illustration of adjustment strategies with (a) direct adaptations and (b) indirect adaptations enabling product substitution strategies. In panel (a), an initial shock to the availability of rice in India, depicted in red and by a bar reaching zero, showing a complete loss of availability. The shock propagates through trade and production links. If for example Bangladesh experiences a substantial reduction in rice availability (blue), as indicated by the lower level of the bar at $t=1$, it triggers the adjustment mechanism to mitigate losses from $t=2$ onwards. Potential beneficial adjustments include: (I) modifying trade links to import the missing product from alternative sources, such as Thailand, (II) adjusting production links to (i) allocate more of the product to self-replicating processes, or (ii) increase output towards the missing item from other processes, and (III) altering the allocation scheme to reduce exports and boost production or consumption. In panel (b), if Bangladesh additionally continues to experience reduced availability of rice at $t=2$, it has the additional option of importing substitute products, here maize, that can be used in similar production processes, from a trade partner, here Australia, thereby mitigating the impact on secondary products, such as live animals, to which both rice and maize can be fed, from $t=3$ onwards. Credit: flags, SVG Repo under a MIT license; country shapes and food icons, SVG Repo under no license}
\end{figure}
\newpage
 
\chapter{Results}\label{Results}

For each sector  $(a,i)$, representing a combination of area $a$ and item $i$, the available amount of product $x_a^i(t)$ at simulation time-step $t$ equals the added production output and trade imports for each iteration of the simulation. A shock to the sector is initiated by setting the production output to $0$ for each time-step of the simulation. We yield the loss $L_{a\rightarrow b}^{i\rightarrow j}$ to a receiving sector $(b,j)$ following a shock to sector $(a,i)$ by comparing the available amount after the shock $x_{a\rightarrow b}^{i\rightarrow j}(t)$ to the baseline scenario $\underline{x}_b^j(t)$ in terms of absolute losses per capita at the end of the simulation $t=\tau$,
\begin{equation}
L_{a\rightarrow b}^{i\rightarrow j}=\frac{\underline{x}_b^j(\tau)-x_{a\rightarrow b}^{i\rightarrow j}(\tau)}{z_b} \quad
\label{absolute loss pc_shock}
\end{equation}
where $z_b$ represents the population of recipient area $b$ in $2020$, as reported by [\hyperlink{29}{29}].\\

Parameters defining the production and trade functions may need adjustment in response to substantial shortages (Figure \ref{model overview} a)). For instance, if a sector $(a,i)$ experiences a substantial loss $l_S$ at time $t=1$ in a simulation run, any parameter related to this sector, here exemplified by production output rate $\alpha^p_{a,i,\text{orig}}$ that defines the output rate of process $p$ towards sector $(a,i)$, is adapted using parameter-specific adaptation matrices. The adjusted output rate $\alpha_{a,i,\text{adap}}^p$ is calculated as follows:
\begin{equation}
\alpha_{a,i,\text{adap}}^p=l_S \cdot {W_\alpha}_{a,i}^p \cdot \alpha_{a,i,\text{orig}}^p + l_S \cdot {R_\alpha}_{a,i}^p \quad
\end{equation}
where $W_{\alpha}$ represents weight adjustment and $R_{\alpha}$ represents rewiring. This adaptation step allows alternating production and trade strategies pertaining the affected item directly.

If compensation efforts in sector $(a,i)$ are insufficient within one time-step, the model allows for increased imports of substitute item $j$ by applying a substitutability factor $S_{ij}^a$. The trade parameter $T_{ab}^j$, which represents the share of exports from sector $(b,j)$ imported into sector $(a,j)$, is adjusted as follows, 
\begin{equation}
    T_{ab,\text{adap}}^j = (1+S_{ij}^a)\cdot T_{ab,\text{orig}}^j \quad.
\end{equation}
A detailed description of the derivation and application of the adaptation matrices $W$ and $R$ for all model parameters, as well as the substitution matrix $S$, can be found in sections \ref{Model calibration} to \ref{Shock adaptation rules}.


\section{Food Availability Shocks in the Real World}\label{Food Availability Shocks in the Real World}
To derive meaningful and stable adjustment rules, we base them on events that had a substantial impact on food availability in a given region. Let ${x_T}_a^i$ denote the available quantity within sector $(a,i)$ in year $T$. For an event to qualify, it must meet three criteria simultaneously: (1) The sector must show a substantial relative decrease $\Delta_{\text{rel}}$ in available quantities within one year, as defined in \eqref{relative loss}; (2) the largest relative decrease must correspond to a substantial absolute decrease $\Delta_{\text{abs}}$ as defined in \eqref{absolute loss}, indicating the shock is significant in both absolute and relative terms; and (3) the event must show low variability $\Delta_{\text{dev}}$ of the amount before and after the drop, as defined in \eqref{deviation}, ensuring the event is well distinguishable from the background noise of the corresponding time series. If the timing of the largest relative drop does not meet the other criteria, the time series does not contain a relevant event.\\

A hyperparameter search combined with a train/test split of the data was performed to identify the set of thresholds that would result in a robust set of adjustment rules, while balancing the number of sectors for which rules can be identified. We found that $\Delta_{\text{rel}}=0.26$, $\Delta_{\text{abs}}=1000$ tonnes and $\Delta_{\text{dev}}=0.32$ is the optimal setup to cover almost the entire network with reasonably stable adaptation rules. For a detailed description of the derivation of the event set, see section \ref{Event definition} and Appendix \ref{Stability of application rules and optimal event limits}.\\

We obtain more than $2,300$ relevant events from $23,616$ time series, covering $120$ of the $123$ food items and $177$ of the $192$ countries. We find that most of the adjustment parameters have a high Matthew correlation coefficient $M$ between their values in the training and test data, ranging from $0.5$ to $0.9$ with item substitutability being the only exception, showing moderate correlation values for both relative limits and deviation limits, ranging from $0.2$ to $0.4$ (see Appendix \ref{Stability of application rules and optimal event limits}).

\section{Adaptation Strategies}\label{Adaptation Strategies}
We consider two classes of adaptation rules: weight adjustment (changing the weight of supply links) and rewiring (creating new supply links). Comparing the number of rules found, weight adjustment is much more common than rewiring (table \ref{adaptation_rule_table}). Existing trade agreements, knowledge on topics such as regulatory environment and business practices and political and diplomatic ties make strengthening existing partnerships less time consuming and costly and therefore more favorable. New trade partnerships usually require significant initial investments e.g. to build transportation capacities and quality controls. Hence, there are far more opportunities for weight adjustments of existing trade relations than to create new ones which we focus on below.\\

We estimate the potential impact $\iota_{ab}$ \eqref{trade_weight_adjustment_impact} of applying an import trade weight adjustment ${W_T}_{ab}$ for a pair of countries $a$ and $b$ as the change in the amount available before and after adjustment, averaged over all relevant products (see Section \ref{Trade adaptation}). The results for the most important rules are shown in Tab. \ref{trade_adaptation_substitutes_table} a). In particular, relations between neighbouring countries intensify in the event of shocks, but size also matters. China, the world's second-largest economy and second-most populous country, shows the potential to gain huge additional trade volumes from other global players such as the United States. Having the highest multiplier does not imply the highest impact. For example, the \textit{Netherlands -- Germany} link has a higher impact than the \textit{Syrian Arab Republic -- Turkey} link, despite having a lower multiplier, because the initial trade value is much higher.\\

\begin{figure}
    \centering
    \vspace{1cm}
\includegraphics[width=\linewidth]{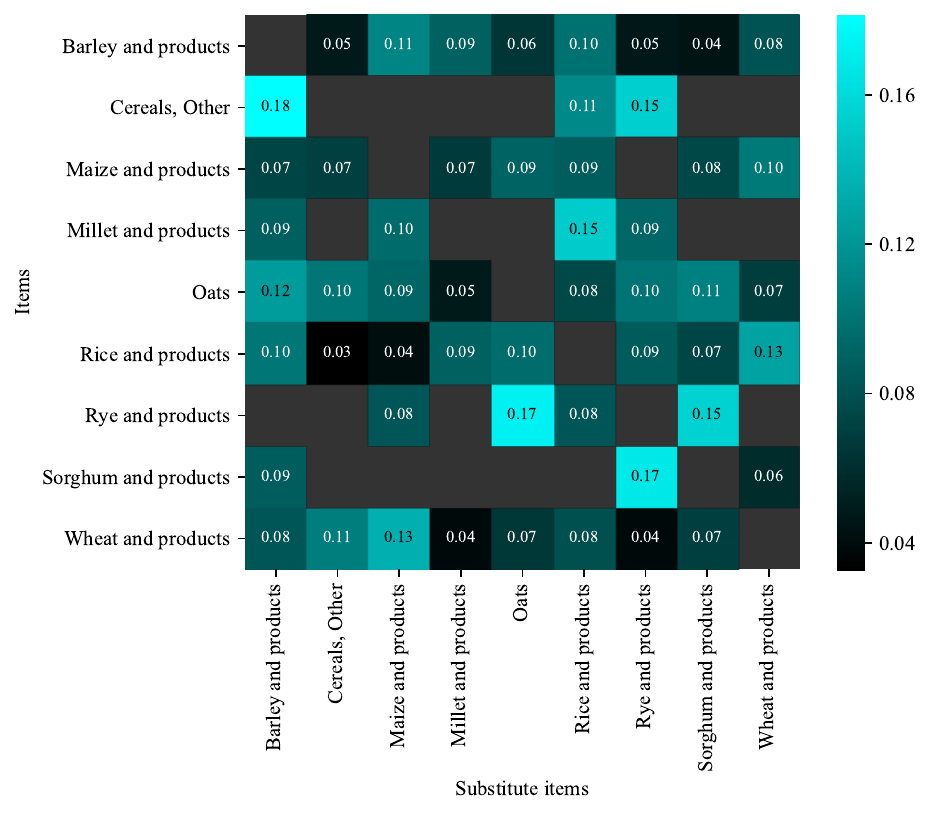}
\caption{Substitution index matrix, $S_{ij}$, for the cereal product group. The entries are the correlation coefficients between the decreased availability (due to an event) of by an item (row) and the increase in imports of its substitution cereal (column) derived according to \ref{Item substitutability}. Not available values are indicated by a grey cell.}
\label{trade_substitute_cereals_figure}
\end{figure}

We allow for product substitution within the same commodity group (see Appendix \ref{Commodity groups}). Here, we illustrate the results for the group of cereals that includes some of the most important basic products such as wheat and maize. These products are used not only for direct consumption but also as inputs in a wide range of processes [\hyperlink{20}{20}]. Figure \ref{trade_substitute_cereals_figure} shows the aggregated substitution matrix $S_{ij}$ for cereals, averaged over all countries. Of the $72$ possible substitutions pairs, $53$ show a significant correlation. In particular, barley, oats, rice and wheat seem to be easily substitutable, as we found substitution relations with all other items in the cereals product group.

Table \ref{trade_adaptation_substitutes_table} b) lists the most impactful substitution relations according to the substitution impact $\iota_{ij}$ \eqref{substitution_impact} (see Methods \ref{Item substitution}). Among the top $10$ relations, wheat and maize are the most frequent substitutes with substitution indices $S_{ij}$ ranging from $0.081$ to $0.127$.

\begin{table}
\caption{Overview of impactful adjustment rules. We show the top $10$ most impactful (a) import trade weight adjustment multipliers and (b) substitution rules. ${W_T}_{ab}$ is the multiplier for the existing trade relations to country $a$ from country $b$ whenever any product in country $a$ experiences a substantial shortage, according to the estimated impact $\iota_{ab}$ (see Section \ref{Trade adaptation}). For the substitution rules, $S_{ij}+1$ is the multiplier applied whenever any country experiences a shortage of food item $i$ and increases its imports of item $j$ as a substitute according to the estimated impact $\iota_{ij}$ (see Section \ref{Item substitution}).} 
    \centering
    \begin{tabular}{ p{0.2cm} | p{4cm} | p{4cm} | C{1cm} | C{4.5 cm} | } 
    \cline{2-5}
    \multirow{2}{*}{\textbf{a)}} & \multirow{2}{*}{\textbf{Importing country $a$}} & \multirow{2}{*}{\textbf{Exporting country $b$}} & \multirow{2}{*}{\boldmath{${W_T}_{ab}$}} & \textbf{Estimated impact} {\boldmath{$\iota_{ab}$}} \\
    & & & & \textbf{in \boldmath{$10^6$} tons} \\ 
    \cline{2-5}
    & Netherlands & Germany & 3.69 & 8.05 \\ 
    & Syrian Arab Republic & Turkey & 13.31 & 5.46 \\ 
    & Guatemala & El Salvador & 8.63 & 3.96 \\
    & China, mainland & United States of America & 9.18 & 3.46 \\
    & Romania & Hungary & 7.34 & 3.45 \\
    & Germany & Netherlands & 2.29 & 1.99 \\
    & Germany & Denmark & 4.47 & 1.83 \\
    & Afghanistan & Pakistan & 1.77 & 1.53 \\
    & China, mainland & Thailand & 5.14 & 1.52 \\
    & China, mainland & Brazil & 1.45 & 1.25 \\
    \cline{2-5}
\end{tabular}
    \vspace{1cm}
\begin{tabular}{ p{0.2cm} | p{4cm} | p{4cm} | C{1cm} | C{4.5 cm} | }
 \cline{2-5}
 \multirow{2}{*}{\textbf{b)}} & \multirow{2}{*}{\textbf{Item $i$}} & \multirow{2}{*}{\textbf{Substitute item $j$}} & \multirow{2}{*}{\boldmath{$S_{ij}$}} & \textbf{Estimated impact in \boldmath{$\iota_{ij}$}}\\ 
  & & & & \textbf{in \boldmath{$10^6$} tons}\\ 
 \cline{2-5}
 \cline{2-5}
 & Rice and products & Wheat and products & $0.127$&$1.15$\\ 
 & Maize and products & Wheat and products & $0.104$&$1.13$\\ 
 & Barley and products & Wheat and products  & $0.081$&$1.11$\\ 
 & Oats & Maize and products & $0.110$&$1.09$\\ 
 & Sorghum and products & Wheat and products & $0.069$&$1.08$\\ 
 & Wheat and products & Maize and products & $0.134$&$1.06$\\ 
 & Barley and products & Maize and products & $0.110$&$1.04$\\ 
 & Millet and products & Maize and products & $0.095$&$1.03$\\ 
 & Oats & Maize and products & $0.092$&$1.02$\\ 
 & Rye and products & Maize and products  & $0.083$&$1.02$\\ 
 \cline{2-5}
\end{tabular}
\vspace{0.5cm}
\label{trade_adaptation_substitutes_table}
\end{table}

\newpage

\section{Model simulations}

\subsection{Adaptation in isolated shock scenarios: India and Ukraine}

We demonstrate the model by considering two isolated shock scenarios: a complete shock to Indian rice exports and Ukrainian wheat exports. For both scenarios, we analyze two conditions: one where the affected sectors can adapt by adjusting trade and production parameters (\textit{India\_adap} and \textit{Ukraine\_adap}), and one where all parameters remain static throughout the simulation (\textit{India\_stat} and \textit{Ukraine\_stat}).\\

\begin{figure}[!htb]
  \includegraphics[width=0.99\linewidth]{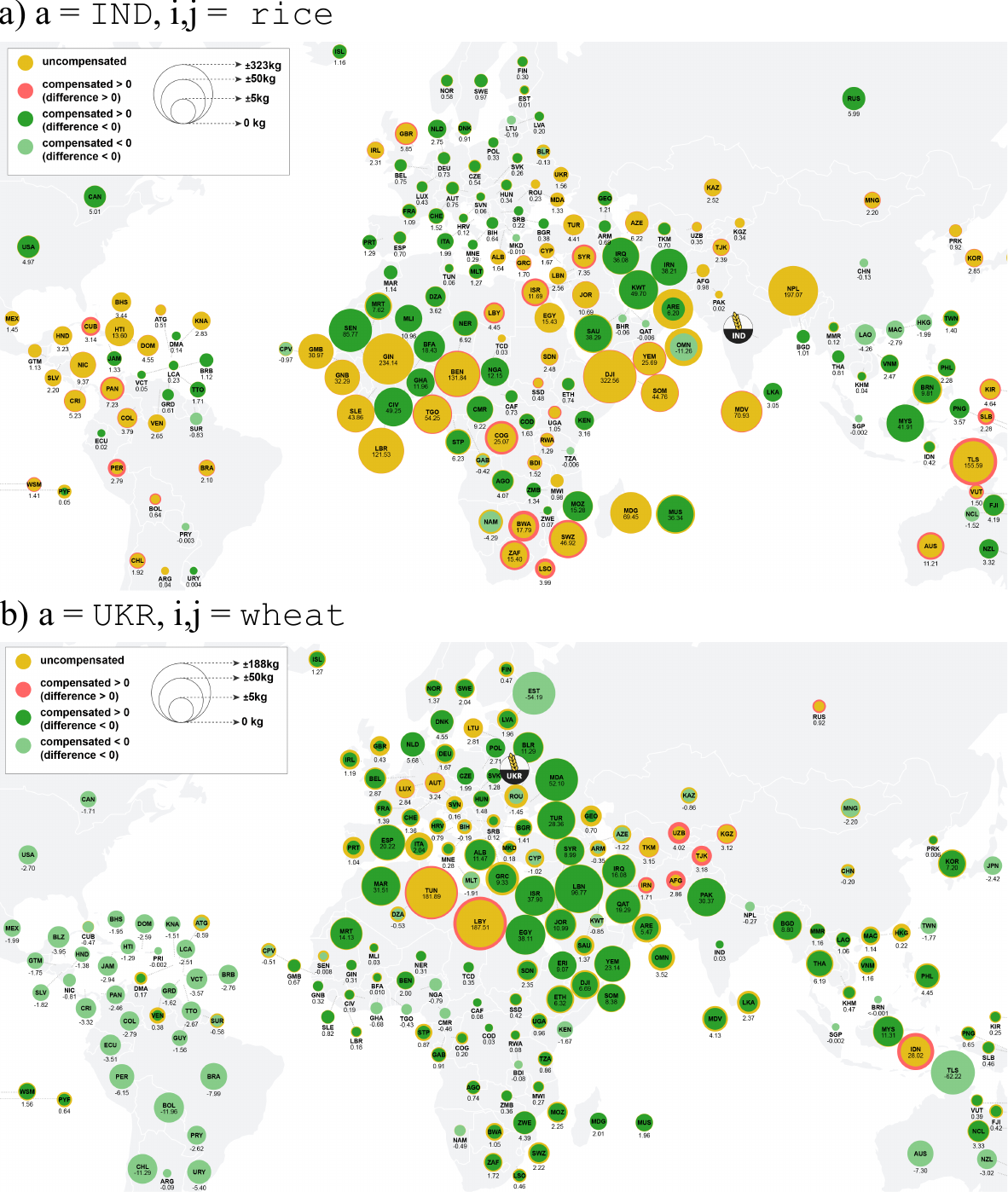}
\caption{Losses in global rice and wheat availability for a $100$ \% shock to (a) Indian rice ($a=\text{\texttt{IND}}$, $i,j=\text{\texttt{Rice and products}}$) and (b) Ukrainian wheat ($a=\text{\texttt{UKR}}$, $i,j=\text{\texttt{Wheat and products}}$) production in kg per capita. Uncompensated losses (yellow circles) ignore a country's adaptive capacity. Some countries experience additional losses (red circles) due to their own or others' compensatory efforts, while others manage to reduce losses (dark green circles) or even achieve a net gain compared to the baseline scenario (light green circle). The size of the circles in corresponds to the amount of losses on a logarithmic scale. The grey arrows in the insets of panel (a) and (b) point towards the outlier of a simultaneous shock to soybeans from Brazil and the United States.}
\label{loss_pc_map}
\end{figure}

\textbf{Indian Rice Scenario:} In the Indian rice scenario ($a=\text{\texttt{IND}}$, $i,j=\text{\texttt{Rice and products}}$), we find that adaptation efforts lead to mixed results for absolute losses per capita $L_{a\rightarrow b}^{i\rightarrow j}$ (Figure \ref{loss_pc_map} a)). Uncompensated losses (yellow circles) ignore a country's adaptive capacity (\textit{India\_stat}). Some countries experience additional losses (red circles) due to their own or others' compensatory efforts, while others manage to reduce losses (dark green circles) or even achieve a net gain compared to the baseline scenario (light green circle) (\textit{India\_adap}). Some of the countries $b$ most affected without compensation, such as the United Arab Emirates, Kuwait, Mauritius, Saudi Arabia and Oman, are able to mitigate some or all of their losses. However, many others, such as Djibouti, Liberia, Benin, Timor-Leste, or the Maldives, fare worse. Djibouti experiences the highest losses, reaching up to $322.6$ kg per person in \textit{India\_adap}. Both Djibouti and the UAE import almost all of their rice, with India providing at least $75$ \% of their supply. The UAE manages to make up some of the shortfall by sourcing additional rice from Vietnam or Thailand, while Djibouti cannot find new suppliers within our framework. In Western Europe, where losses are small, most countries seem to benefit slightly, while other regions show mixed results. In fact, countries with a Human Development Index (HDI) below $0.6$ [\hyperlink{30}{30}] show, on average, $4.8$ \% higher losses in the scenario \textit{India\_adap} compared to \textit{India\_stat}, while countries with a HDI above $0.8$ experience $14.2$ \% less losses. Overall, compensatory efforts did only mitigate about $1$ \% of global rice losses, but about $30$ \% of the world's rice supply is irretrievably lost due to the shock.\\

\textbf{Ukrainian Wheat Scenario:} The positive effects of adaptation are clearly seen in the Ukrainian wheat scenario ($a=\text{\texttt{UKR}}$, $i,j=\text{\texttt{Wheat and products}}$), for which the non-adaptation baseline (here \textit{Ukraine\_stat}) has already been studied [\hyperlink{20}{20}]. We find that adaptation efforts generally lead to lower losses. As shown in Fig. \ref{loss_pc_map} b), most of the affected countries show reduced losses in \text{Ukraine\_adap}, except for Tunisia, Libya, and Indonesia. Lebanon, initially experiencing higher losses than Libya, manages to reduce the impact despite having a high dependence on Ukraine for about $71$ \% of its wheat imports. Compared to the Indian rice shock, which affects up to $96$ \% of a single country's rice imports, the Ukrainian wheat shock has a smaller impact. Many countries experience positive spillover effects and even net gains compared to the unshocked baseline.\\

\subsection{Adaptation in multiple crises: combined shocks}
To assess the impact of two simultaneous shocks to sectors $(a_1,i_1)$ and $(a_2,i_2)$ on a receiving sector ($b$,$j$), we calculate the shock superposition impact $SI_{a_1,a_2\rightarrow b}^{i_1,i_2\rightarrow j}$. This allows to conclude whether the shock superposition has a sub-additive, super-additive or no effect on the losses. This is done by comparing the combined losses from the superposed scenario $L_{a_1,a_2\rightarrow b}^{i_1,i_2\rightarrow j}$ with the sum of losses from individual shock scenarios as follows,
\begin{equation}
    SI_{a_1,a_2\rightarrow b}^{i_1,i_2\rightarrow j} = L_{a_1,a_2\rightarrow b}^{i_1,i_2\rightarrow j}-(L_{a_1\rightarrow b}^{i_1\rightarrow j} +L_{a_2\rightarrow b}^{i_2\rightarrow j}) \quad .
    \label{superposition_impact}
\end{equation}
If $SI_{a_1,a_2\rightarrow b}^{i_1,i_2\rightarrow j} > 0$, the losses indicate a super-additive shock effect, $SI_{a_1,a_2\rightarrow b}^{i_1,i_2\rightarrow j} < 0$ a sub-additive impact.\\

We now simulate the combined shocks of Indian rice ($a_1=\text{\texttt{IND}}$, $i_1=\text{\texttt{Rice and products}}$) and Ukrainian wheat ($a_2=\text{\texttt{UKR}}$, $i_2=\text{\texttt{Wheat and products}}$) and compare the superposition impact $SI_{a_1,a_2\rightarrow b}^{i_1,i_2\rightarrow j}$ \eqref{superposition_impact} for this scenario with and without adjustment.\\

\textbf{Combined Shock Scenario Without Adaptation:} In this scenario, we find that the losses are numerically identical to the sum of in the isolated wheat and rice shocks for any receiving sector $(b,j)$, leaving the superposition impact $0$. This indicates no interaction between the two shock propagation processes.
\\
 
\textbf{Combined Shock Scenario With Adaptation:} When adaptive dynamics are considered, the results change. Focusing on the superposition impact $SI_{a_1,a_2\rightarrow D}^{i_1,i_2\rightarrow K}$ \eqref{superposition_impact_general} on directly affected commodities globally as with $D=\mathcal{A}$ (all countries covered by the model) and $K = $ \texttt{Rice and products, Wheat and products}, the combined shocks have a sub-additive effect of $SI_{a_1,a_2\rightarrow D}^{i_1,i_2\rightarrow K}=-0.05$ kg per capita, meaning $0.05$ kg more rice and wheat are available per person in the superposed shock scenario.

However, when considering superposition impact on other food categories, the results are fundamentally different. Examining $SI_{a_1,a_2\rightarrow D}^{i_1,i_2\rightarrow K}$ \eqref{superposition_impact_general} across the whole network with $D=\mathcal{A}$ and $K = \mathcal{I}$ (all items covered by the model), the result is $SI_{a_1,a_2\rightarrow D}^{i_1,i_2\rightarrow K} = 14.09$ kg losses per capita. Super-additive effects emerge. The $14.09$ kg additional losses per capita represent about $0.1$\% of the average available volume per capita in relative terms.\\

The combined rice and wheat shocks are of particular interest due to their geopolitical relevance as rice and wheat are direct substitutes. However, the question remains whether the sub-additive effect on the shocked items and the super-additive effect on the whole network persists when initiating other pair of shocks or whether it is only an individual case of this specific pair. We therefore investigate the superposition effect of $1,000$ superposed shocks, each randomly selecting two sectors $({a_r}_1,{i_r}_1)$ and $({a_r}_2,{i_r}_2)$, sampled from the $100$ largest exporting primary products sectors by production volume (see \ref{Evaluating effects of combined shocks}).\\
 
\begin{figure}[!htb]
\vspace{1cm}
\includegraphics[width=0.9\linewidth, valign=t]{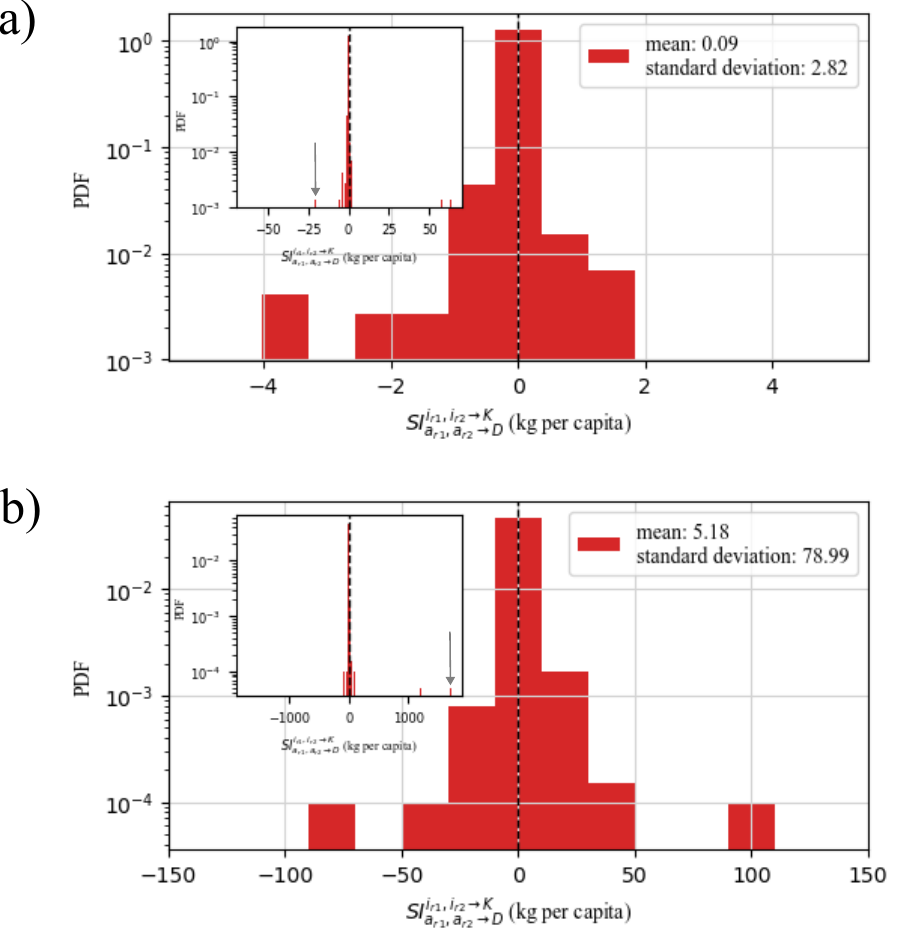}
\vspace{1cm}
\caption{Histograms of $SI_{{a_r}_1,{a_r}_2\rightarrow D}^{{i_r}_1,{i_r}_2\rightarrow K}$ for sampled $({a_r}_1, {i_r}_1)$ and $({a_r}_2, {i_r}_2)$. Results are shown for (a) the shocked items ($D=\mathcal{A}$ and $K=\{{i_r}_1,{i_r}_2\}$) and (b) summed over all food items ($D=\mathcal{A}$ and $K=\mathcal{I}$) as probability density functions. Both distributions centre around zero (black dotted line), with some outliers shown in the respective insets. The distribution of the effect on the shocked items themselves shows no evidence of sub- or super-additivity. However, the aggregated network effect indicates a slight super-additivity (p-value: $0.038$). The grey arrow indicate one example of an extreme outlier in terms of superposition impact, a simultaneous shock to soybean production in the United States and Brazil.}
\label{superposition_histogram}
\vspace{1cm}
\end{figure}

\textbf{Superimposed Sampled Shock Scenarios With Adaptation:} Similar to the case discussed above the superimposed scenarios of the two random shocks show super-additive effects on the whole network. Fig. \ref{superposition_histogram} a) shows the distribution of $SI_{{a_r}_1,{a_r}_2\rightarrow D}^{{i_r}_1,{i_r}_2\rightarrow K}$ with $D=\mathcal{A}$ and $K=\{{i_r}_1,{i_r}_2\}$, representing global superposition impact in products of the two shocked sectors. The distribution centers around $0$, with a mean $\langle SI_{{a_r}_1,{a_r}_2\rightarrow D}^{{i_r}_1,{i_r}_2\rightarrow K}\rangle_{{i_r}_1,{i_r}_2} = 0.09$ kg per capita. Fig. \ref{superposition_histogram} b) shows the distribution of $SI_{{a_r}_1,{a_r}_2\rightarrow D}^{{i_r}_1,{i_r}_2\rightarrow K}$ with $D=\mathcal{A}$ and $K=\mathcal{I}$, representing the superposition impact across the entire network, with a mean $\langle SI_{{a_r}_1,{a_r}_2\rightarrow D}^{{i_r}_1,{i_r}_2\rightarrow K}\rangle_{{i_r}_1,{i_r}_2} = 5.18$ kg per capita. The null-hypothesis that the mean is $0$ can be rejected with a p-value of $0.038$, clearly indicating a super-additive impact of superposition of shocks on losses of the entire network. 

Fig. \ref{superposition_histogram} a) and \ref{superposition_histogram} b) include insets showing scenarios with extreme outliers in additional or reduced losses due to superposition. One notable outlier is a simultaneous shock to soybeans from Brazil and the United States (grey arrow), resulting in $SI_{a_1,a_2\rightarrow D}^{i_1,i_2\rightarrow K}=-20$ kg per capita for soybeans and $SI_{a_1,a_2\rightarrow D}^{i_1,i_2\rightarrow K}=1717$ kg per capita for all items. For comparison, a $20$ kg per capita reduction in soybean losses corresponds to a relative gain of $40$ \%, and a $1717$ kg per capita increase in losses across the network corresponds to an additional relative loss of $12$ \%.

\newpage

\newpage
\chapter{Discussion}\label{Discussion}
 
We conducted a comprehensive study of virtually all food availability shocks over $35$ years and documented the responses to these shortages. Our findings indicate that the vast majority of shock responses involved volume adjustments to existing suppliers, and to a less frequent degree searching for and establishing new supply possibilities. By, quantifying these responses, we were able to calibrate a model that estimates actual food shortages following specific events, explicitly considering national response options. The adaptation rules derived from historical events in extensive food IO data [\hyperlink{19}{19}] were applied to a multilayer network model of food production and trade [\hyperlink{20}{20}, enabling countries experiencing shortages in any of the $123$ included products to adjust their trade and production strategies to mitigate the impact of incoming shocks. 

Based on $2,300$ relevant events, we demonstrated that stable and consistent trade and production adjustment strategies can be inferred from the data on a country- and product-specific basis. In terms of trade adjustment, relations between neighbouring countries and major economic players appear to have the largest potential impact in absolute terms. The initial strength of the adjusted trade relationship proves to be highly substantial, as a high multiplier does not necessarily lead to the highest additional trade volumes. Substitutions, which we only allow for items from the same product group, are particularly relevant for cereals. Cereals are a crucial product group since they are not only consumed directly but also enter many production processes.\\

We further illustrated the adaptive capabilities of our model by simulating two shock scenario: one affecting Indian rice [\hyperlink{3}{3},\hyperlink{4}{4}] and one affecting Ukrainian wheat [\hyperlink{5}{5},\hyperlink{6}{6}] production. In \textit{Indian\_adap}, compensation strategies were at least partially successful, mitigating losses in agricultural products in some areas but causing both negative and positive spillovers elsewhere. Globally, the loss of Indian rice could not be fully compensated, with only $1$ \% of the losses being saved, highlighting the strategy's limitation in merely redistributing the remaining available products.

The large impact of adaptation on the modeled shocks on individual countries may seem surprising, given the relatively low complexity of food products [\hyperlink{31}{31}]. However, countries most affected by the shocks tend to have high import concentrations, making them highly vulnerable. Indian rice can account for up to $96$ \% of a country's rice imports. Diversification is an obvious remedy, but building alternative strategies requires effort especially in the case of rice [\hyperlink{32}{32}]. The observed compensation processes in response to reduced rice supplies disadvantage economically weaker countries. Countries with a low HDI experienced on average $4.8$ \% more losses after compensation compared to the uncompensated scenario, while countries with a very high HDI experienced $14.2$ \% fewer losses. 

In the case of \textit{Ukraine\_adap}, the compensation adjustments of trade and production led to much lower total wheat losses across the global food network, with almost every affected country managing to increase wheat availability compared to \textit{Ukraine\_stat}. The differences in the overall success of the compensation strategies may be explained by the size of the shock relative to the global supply and the degree of diversification of the importing countries. India's significant share in the global rice market means that some countries are highly dependent on imports from them. Conversely, the global supply of wheat is more diversified, and countries do not rely as heavily on wheat imports from Ukraine. This difference in reliance and supply diversification likely contributes to why compensation strategies are more successful in the Indian rice scenario compared to the Ukrainian wheat scenario.

We next studied how multiple simultaneous crises affect global food security by examining a hypothetical scenario of combined Ukrainian wheat and Indian rice shocks. Without adaptation, the combined shock's effect equaled the sum of the isolated effects rice and wheat shocks. However, with adaptive capacity considered, we observed sub-additive losses for rice and wheat. We find super-additivity in losses when looking at the cumulative impact across all regions and food products. Sampling combined shocks delivers a similar result for the superposition impact across the entire network. We find combinations of shocks that are especially systemically risky, such as a simultaneous shock to soybeans in Brazil and the US. Superposing these two shocks leads to highly exacerbated losses of additional $12$ \% available volume lost across all region and food products.\\

The model has several limitations. First, it provides a simplified representation of an adaptive food supply network, allowing for isolated, limited adjustments in response to severe losses. The model is designed to adjust its parameters only twice, rather than continuously, in response to indirect subsequent losses. Furthermore, it does not allow switching between production processes, as this would require taking into account the capacity of production facilities [\hyperlink{33}{33}], which is currently neglected.

Second, the model only considers linear production functions, whereas implementing Leontief production functions would provide a more comprehensive representation of actual production outputs and critical production inputs [\hyperlink{34}{34}]. Additionally, the model does not account for non-agricultural inputs or outputs, including critical inputs such as fertilisers [\hyperlink{35}{35}], which play a crucial role in agricultural productivity.

Third, the model does not consider the stocks that countries could draw on in the event of shocks and which could also be depleted [\hyperlink{21}{21},\hyperlink{36}{36}]. Currently, the available product quantity is derived solely from trade and production. Although a portion of the yield can be allocated towards production, this effectively removes it from further usage.

Fourth, the granularity of the model is constrained by the resolution of the FABIO database. More detailed spatial data at a subnational level could offer deeper insights into which exact regions within a country are more severely affected by shocks [\hyperlink{37}{37}]. This would enable the modeling of more realistic shock scenarios, as disruptions in agricultural supply, especially those caused by climate catastrophes, do not uniformly affect entire countries but can impact various subnational regions differently. Additionally, the granularity of product categories in the database presents challenges in accurately assessing the availability of specific food items [\hyperlink{34}{34}]. For example, the category labeled 'Wheat and products' encompasses not just wheat but also wheat-derived products such as flour, bread, and pasta. The actual composition may even vary between countries due to differing reporting practices.

Fifth, shock propagation and adjustment mechanisms are modelled in the absence of prices and markets, which typically play a crucial role in supply dynamics [\hyperlink{9}{9},\hyperlink{21}{21}] as competition increases when a product supply diminishes, but demand remains constant. 

Sixth, and finally, the shocks considered are relatively simple compared to real-world complexities. For example, the Indian rice embargo only partially bans the export of certain types of rice, while imposing price barriers and export restrictions on others. The use of a complete shock to all rice production in India represents a simplification and can be seen as a worst-case scenario.\\

Despite its limitations, this paper substantially enhances our understanding of how adaptive strategies can mitigate the impact of shocks on global food production systems. By integrating historical sector-specific data-driven adaptation rules into a multilayer network model, we can make more accurate predictions of food availability following specific events. This approach provides a valuable framework for policymakers and researchers to develop more effective strategies for ensuring food security, especially in the face of multiple, simultaneous crises.

The study highlights the importance of trade relationships, the need for diversification, and the potential for increased inequality following food shortages. Future research should focus on incorporating more detailed production processes, non-linear production functions, and market dynamics to enhance the model's realism. Additionally, exploring the role of stockpiles and a refined spatial resolution will provide deeper insights into regional vulnerabilities and adaptive capacities.

Ultimately, this research underscores the critical role of adaptive capacity in global food security and lays the groundwork for developing more robust strategies to mitigate the impacts of global food shortages. This work highlights the ongoing need to build resilient food supply systems that can effectively respond to and recover from diverse and complex crises.
\newpage
\chapter{Methods}\label{Methods}

\section{Model calibration}\label{Model calibration}

The FABIO database [\hyperlink{19}{19}] spans $35$ years, and for each of these years, the parameters of the static model can be calculated as described in [\hyperlink{20}{20}]. Our approach relies on comparing the parameters of two consecutive years. Consequently, we exclude any country that did not exist throughout the entire calibration period to avoid generating incorrect transition parameters due to non-existent data for some years. The dataset starts in $1986$, but we begin our analysis in $1992$ to account for the significant geopolitical changes in $1991$, particularly the dissolution of the Soviet Union and the emergence of $15$ new states, including major agricultural players like Russia and Ukraine.\\

The model parameters $\alpha_{a,i}^p$, $\beta_{a,i}^p$, $\nu_{a,i}^p$, $T_{ab}^i$, $\eta_{a,i}^{\text{exp}}$ and $\eta_{a, i}^{\text{prod}}$ and the initial vector ${x_0}_a^i$ as detailed in [\hyperlink{20}{20}] are derived from the $2020$ FABIO data. The $2020$ FABIO dataset comprises $192$ countries (\textit{areas}), denoted as $\mathcal{A}$, and a set of $123$ agricultural products (\textit{items}), denoted as $\mathcal{I}$, produced by $117$ production processes (\textit{processes}), denoted as $\mathcal{P}$. The exact derivation process is elaborated in [\hyperlink{20}{20}]. The unique 2-tuple $(a,i)$ of a country $a$ and an item $i$ is referred to as a sector, resulting in a total of $23,616$ sectors.\\

To avoid unrealistic adjustments, we set a threshold of $10^{-3}$ for parameters that denote shares and sometimes have extremely low values (e.g., $10^{-90}$ or lower). Parameters below this threshold are set to zero to ensure the model creates new links rather than making disproportionately strong adjustments to negligible links.

\section{Model without compensation}\label{Model without compensation}
We use a recent bipartite multilayer model for the propagation of food supply shocks [\hyperlink{20}{20}] as our basic model. This model (figure \ref{model}) quantifies the available amount of a product in a country that is obtained through trade or production through an iterative three-step process: production, trade and allocation. This process is repeated at each time step of the simulation. During production, input products are transformed into output products through production processes; during trade, products are traded with other countries; and finally, during allocation, all products previously obtained through production or trade are allocated to different purposes for which they will be used in the next time step.\\

The model allows for simulating regional production shocks and observing direct and indirect losses in all (other) products and regions. The impact of shocks is measured by comparing the amount available in a shock scenario with the baseline case without shocks.\\

The output of an item $i$ in an area $a$, in hereafter referred to as sector $(a,i)$, is quantified by summing the outputs of each production process $p$ that produces the item. $\alpha_{a,i}^p$ represents the output rate from processes $p$ towards sector $(a,i)$ that require input items according to the model, while $\beta_{a,i}^p$ covers processes that can produce without food inputs. Either $\alpha_{a,i}^p$ or $\beta_{a,i}^p$ can be greater than zero for one combination of $a$,$i$ and $p$, but not both. Processes with $\beta_{a,i}^p > 0$ typically refer to the harvesting of primary agricultural products where seeds are not recorded as inputs. The input share for process $p$ from the for-production allocated share of sector $(a,j)$ $p_a^j(t-1)$  is denoted by $\nu_{a,j}^p$, resulting in output $o_a^i(t)$,
\begin{equation}
o_a^i(t)=\sum_{p \in \mathcal{P}} (\alpha_{a,i}^p 
\sum_{j \in \mathcal{I}}(\nu_{a,j}^p p_a^j(t-1)) + \beta_{a,i}^p) \quad.
\label{output}
\end{equation}
The amount obtained from trade $h_a^i(t)$ is calculated by summing the imports from trading partners from which item $i$ is imported. $T_{ab}^i$ indicates the shares of the for-trade reserved volumes $e_b^i(t-1)$ of exporting sector $(b,i)$ that are to be received by country $a$,
\begin{equation}
h_a^i(t)=\sum_{b\in \mathcal{A}} T_{ab}^i e_b^i(t-1) \quad.
\label{trade}
\end{equation}
At the end of a time step $t$, sector $(a,i)$ has a total amount $x_a^i(t)$ available,
\begin{equation}
x_a^i(t)=o_a^i(t)+h_a^i(t) 
\label{amount}
\quad .
\end{equation}
$p_a^i(t)$ and $e_a^i(t)$ represent the shares of $x_a^i(t)$ allocated to production and exports, respectively, according to the individual shares $\eta$ of each sector.
\begin{align}
    \;p_a^i(t) = \eta_{a,i}^{\text{prod}} \cdot x_a^i(t) \quad,\\
    \;e_a^i(t) = \eta_{a,i}^{\text{exp}} \cdot x_a^i(t) \quad.
\end{align}
A shock to the system is induced by artificially destroying a fraction $\phi$ or all ($\phi = 100 \%$) of the output of item $i$ in area $a$ at time step $t$, starting at $t=0$ until the end of the simulation $t=\tau$,
\begin{equation}
o_a^i (t) = (1 - \phi) \cdot o_a^i (t)  \quad \forall t \in \{0, ... , \tau\} \quad.
\label{shock initiation}
\end{equation}
The shock propagates through the production and trade network. The model captures not only direct losses due to first-order trade relationships of the shocked area and its direct customers, but also quantifies indirect losses resulting from higher-order trade relationships (Figure \ref{shock_propagation}). It also covers the indirect effects on secondary items that use the initially shocked item upstream in their production. The loss $L_{a\rightarrow b}^{i\rightarrow j}(t)$ of a sector $(b,j)$ due to a shock to sector $(a,i)$ at time $t$ is assessed in comparison to the baseline, which is a scenario $\underline{x}$ of the simulation in which no shock was issued, as an absolute loss per capita,
\begin{equation}
L_{a\rightarrow b}^{i\rightarrow j}(t)=\frac{\underline{x}_b^j(t)-x_{a\rightarrow b}^{i\rightarrow j}(t)}{z_b} \quad.
\end{equation}
$x_{a\rightarrow b}^{i\rightarrow j}(t)$ is the available amount in sector $(b,j)$ after a shock to sector $(a,i)$, while $\underline{x}_b^j(t)$ is the availability in sector $(b,j)$ in the unshocked scenario. $z_b$ is the population of recipient country $b$ in the year $2020$, as reported by [\hyperlink{29}{29}]. We will commonly use $L_{a\rightarrow b}^{i\rightarrow j}=L_{a\rightarrow b}^{i\rightarrow j}(\tau)$ with $\tau=10$ being the length of a simulation run and $L_{a\rightarrow b}^{i\rightarrow j}$ representing the final absolute loss per capita, which serve as the indicator of a shock's severity.

\section{Possible Compensation Dynamics}\label{Trade and production adaptation strategies}
The uncompensated model describes the food trade and production network as an iterative three-step process. A set of parameters specifies the function that, for each step, quantifies the values for each sector of the availability vector. By applying the multipliers or rewiring components to each parameter, a country can implement a variety of strategies:
\begin{itemize}
    \item Increase the volume of imports by strengthening existing or establishing new trade relationships in the vulnerable product.
    \item Adjust the allocation of production inputs, for example by directing more of the compromised product to their own production as seed or regular input (i.e. grain as input for bread versus grain as input for more grain production).
    \item Increase production of the vulnerable product by shifting production rates towards it.
    \item Shift allocation from trade to domestic uses such as production or consumption.
    \item Increase imports of suitable substitutes.
\end{itemize}
\section{Shock adaptation rules}\label{Shock adaptation rules}
\subsection{Growth adjustment}\label{Groth adjustment}
As the focus is on the adaptive behaviour of countries by changing their trade and production strategy, the parameters obtained from the FABIO data are adjusted for the effects of economic growth. Since the parameters $\nu_{a,i}^p$, $T_{ab}^i$, $\eta_{a,i}^{\text{exp}}$ and $\eta_{a,i}^{\text{prod}}$ denote shares and cannot lead to volume increases that would indicate economic growth, they do not need to be adjusted. $\alpha_{a,i}^p$ and $\beta_{a,i}^p$, the output rates of production processes with or without quantifiable inputs, are expressed in absolute terms and can therefore reflect economic growth. To eliminate this effect, we ensure that the output volume of a process $p$ is constant across all its output sectors $(a,i)$. As a reference value, the parameters $\alpha_{a,i}^p$ and $\beta_{a,i}^p$, calculated for the starting year $1992$, are used. We calculate the normalised version $\alpha_{a,i, \text{norm}}^p (T)$ and $\beta_{a, i, \text{norm}}^p (T)$ from the unnormalized original parameters $\alpha_{a,i, \text{unnorm}}^p (T)$ and $\beta_{a,i, \text{unnorm}}^p (T)$ for each year $T$ as follows,
\begin{equation}
\alpha_{a,i, \text{norm}}^p (T)  = \alpha_{a,i, \text{unnorm}}^p (T)\cdot \frac{\displaystyle\sum_{i \in \mathcal{I}}{\alpha_{a,i, \text{unnorm}}^p (1992)}}{\displaystyle\sum_{i \in \mathcal{I}}{\alpha_{a,i, \text{unnorm}}^p (T)}} \quad,
\label{production_output_alpha_norm}
\end{equation}
\begin{equation}
\beta_{a,i,\text{norm}}^p (T)  = \beta_{a,i, \text{unnorm}}^p (T)\cdot \frac{\displaystyle\sum_{i \in \mathcal{I}} \beta_{a,i, \text{unnorm}}^p (1992)}{\displaystyle\sum_{i \in \mathcal{I}} \beta_{a,i, \text{unnorm}}^p (T)} \quad.
\label{production_output_beta_norm}
\end{equation}
In short, we will use $\alpha_{a,i}^p (T)=\alpha_{a,i, \text{norm}}^p (T)$ and $\beta_{a,i}^p (T)=\beta_{a,i, \text{norm}}^p (T)$ from now on.\\

The initial vector ${x_{0}}_a^i$ is also rescaled against the total amount of globally available items in $1992$ as the absolute available amount can also reflect economic growth. The rescaled initial vector ${x_{0}}_{a,\text{norm}}^i(T)$ is calculated from the original initial vector ${x_{0}}_{a,\text{unnorm}}^i(T)$ for each year $T$ in the calibration, by keeping the total amount of available products across all sector constant,
\begin{equation}
{x_{0}}_{a,\text{norm}}^i(T)  = {x_{0}}_{a,\text{unnorm}}^i(T) \cdot \frac{\displaystyle\sum_{(a,i)\in \mathcal{A}\times\mathcal{I}} {x_{0}}_{a,\text{unnorm}}^i(1992)}{\displaystyle\sum_{(a,i)\in \mathcal{A}\times\mathcal{I}} {x_{0}}_{a,\text{unnorm}}^i(T) } \quad.
\label{x_0_norm}
\end{equation}
For brevity, we set ${x_T}_a^i={x_{0}}_{a,\text{norm}}^i(T)$.

\subsection{Event definition}\label{Event definition}
Units that lost a substantial amount of available goods from one year to another are identified as follows. Where $(a_E,i_E)$ is the event sector, $T_E$ is the time of an event and $l_E$ is the relative loss of available quantity of the sector at $T=T_E$. To be classified as an event, the 4-tuple ($a_E$,$i_E$,$T_E$,$l_E$) must fulfil three criteria,
\begin{itemize}
    \item substantial relative loss of available amount $\Delta_{rel}^{a_E,i_E,T_E}$ exceeding the threshold limit $\Delta_{\text{rel}}$,
\begin{equation}
\Delta_{\text{rel}}^{a_E,i_E,T_E}=\abs{\frac{{x_{T_E}}_{a_E}^{i_E}-{x_{(T_E-1)}}_{a_E}^{i_E}} {{x_{(T_E-1)}}_{a_E}^{i_E}}} >\Delta_{\text{rel}}
    \label{relative loss}
    \end{equation}\\
   with $\Delta_{\text{rel}}^{a_E,i_E,T_E}=l_E$.
    \\
    \item substantial absolute loss of available amount $\Delta_{\text{abs}}^{a_E,i_E,T_E}$ exceeding the threshold limit $\Delta_{abs}$,
    \begin{equation}
    \Delta_{\text{abs}}^{a_E,i_E,T_E}=\abs{{x_{T_E}}_{a_E}^{i_E}-{x_{(T_E-1)}}_{a_E}^{i_E}} > \Delta_{\text{abs}} \quad.
    \label{absolute loss}
    \end{equation}\\
    \item Stable conditions before and after the event. To quantify stability, the coefficient of variation ($CV$) is used and must fall below the threshold $\Delta_{\text{dev}}$,
    \begin{equation}
    CV=\frac{\sigma}{\mu} < \Delta_{\text{dev}}
    \label{deviation}
    \end{equation}
    \begin{equation}
    \text{with} \quad \mu=\langle {x_{T}}_{a_E}^{i_E} \rangle_{T} \quad \text{and} \quad \sigma=\sqrt{\langle {(x_{T}}_{a_E}^{i_E})^2 \rangle_{T} - \mu^2}\quad.
    \end{equation}\\
    If the $CV$ for either $T < T_E$ or $T > T_E$ exceeds the threshold $\Delta_{\text{dev}}$, the event ($a_E$,$i_E$,$T_E$,$l_E$) will not be considered as relevant. To apply this criterion reasonably, potential events with $T_E<1997$ or $T_E>2015$ will not be considered, as there will be too few years before or after the event to quantify stability. We require at least $5$ years to evaluate this criterion. Furthermore, because of this aspect, each starting entry time series can only contain one potential event, otherwise the rest of the time series couldn't be stable.
\end{itemize}

\subsection{Calibration of adaptation rules}\label{Calibration of adaptation rules}
Changes in each of the model parameters $\alpha_{a,i}^p$, $\beta_{a,i}^p$, $\nu_{a,i}^p$, $T_{ab}^i$, $\eta_{a,i}^{\text{exp}}$ and $\eta_{a,i}^{\text{prod}}$ after a relevant event identified in \ref{Event definition} are observed to derive general adaptation rules. The framework of possible post-event adjustments for an example parameter entry $v$ describes the change with a weight adjustment component $w$ and a rewiring component $r$. In the following, $l_E$ represents the loss of an exemplary event $E$. We have 
\begin{equation}
    v(T_E+1)=l_E \cdot w \cdot v(T_E) + l_E \cdot r \quad.
\label{transformation derivation}
\end{equation}
The weight adjustment and the rewiring component can be calculated as follows, since $l_E$ and the variable $v$ before and after the event are known,
\begin{align}
    w &= \left\{
    \begin{aligned}
        &\frac{1}{l_E} \frac{v(T_E+1)}{v(T_E)}, && \text{if } v(T_E) > 0\\
        &0, && \text{if } v(T_E) = 0
    \end{aligned} \right. \quad,\\
    r &= \left\{
    \begin{aligned}
        &0, && \text{if } v(T_E) > 0\\
        &\frac{1}{l_E} v(T_E+1), && \text{if } v(T_E) = 0
    \end{aligned} \right. \quad.
\end{align}
For an event ($a_E$,$i_E$,$T_E$,$l_E$), the observed changes in a to that event related entry of the production output rate $\alpha$ results in the following adjustment components,
\begin{align}
    {w_{\alpha}}_{a_E,i_E}^p &= \left\{
    \begin{aligned}
        &\frac{1}{l_E} \frac{{\alpha}_{a_E,i_E}^p(T_E+1)}{{\alpha}_{a_E,i_E}^p(T_E)}, && \text{if } {\alpha}_{a_E,i_E}^p(T_E) > 0\\
        &0, && \text{if } {\alpha}_{a_E,i_E}^p(T_E) = 0
    \end{aligned} \right. \quad ,  \label{alpha_adaptation_derivation}\\  
    {r_{\alpha}}_{a_E,i_E}^p &= \left\{
    \begin{aligned}
        &0, && \text{if } {\alpha}_{a_E,i_E}^p(T_E) > 0\\
        &\frac{1}{l_E} {\alpha}_{a_E,i_E}^p(T_E+1), && \text{if } {\alpha}_{a_E,i_E}^p(T_E) = 0
    \end{aligned} \right. \quad .
\end{align}
For each individual combination of event sector $(a_e,i_e)$, a specific weight adaptation ${m_{\alpha}}_{a_E,i_E}^p$ and rewiring adaptation ${r_{\alpha}}_{a_E,i_E}^p$ can be determined. For increased robustness and coverage, adaptation rules are inferred for collections of events by averaging over all countries $a_E$ for which a substantial event has been observed for item $i_E$,
\begin{align}
    {W_{\alpha}}_{i_E}^p &=\langle{w_{\alpha}}_{a_E,i_E}^p\rangle_{a_E} &\Rightarrow&\quad {W_{\alpha}}_{i_E}^p = {W_{\alpha}}_{a,i_E}^p  &\forall a \in \mathcal{A} \quad,\\
{R_{\alpha}}_{i_E}^p&=\langle{r_{\alpha}}_{a_E,i_E}^p\rangle_{a_E} &\Rightarrow& \quad {R_{\alpha}}_{i_E}^p = {R_{\alpha}}_{a,i_E}^p  &\forall a \in \mathcal{A} \quad.
\label{alpha_adaptation_aggregation}
\end{align}

For $\beta$ and $\nu$, the weight adjustment and the rewiring transformation matrix can be derived in the same way. The allocation parameters are also derived in a similar way, as the values are aggregated to obtain a full vector, assuming similar adaptive behaviour for items across countries. The trade parameter, on the other hand, describes country-country, not item-process, relationships and therefore averaging over event countries will eliminate relevant information. When completing the transformation parameter for the trade parameter in the direction of imports, the aggregation over all items is carried out by averaging over all items $i_E$ for which a substantial event has been observed in area $a_E$,
\begin{align}
    {W_T}_{a_Eb} &=\langle{w_T}_{a_Eb}^{i_E}\rangle_{i_E} &\Rightarrow&\quad {W_T}_{a_Eb} = {W_T}_{a_Eb}^i  &\forall i \in \mathcal{I} \quad,\\
    {R_T}_{a_Eb} &=\langle{r_T}_{a_Eb}^{i_E}\rangle_{i_E} &\Rightarrow&\quad {R_T}_{a_Eb} = {R_T}_{a_Eb}^i  &\forall i \in \mathcal{I} \quad.
\label{trade transition aggregation}
\end{align}

The export direction is derived in a similar way, but instead of changes from $T_{a_Eb}(T_E)$ to $T_{a_Eb}(T_E+1)$, we observe changes from $T_{ba_E}(T_E)$ to $T_{ba_E}(T_E+1)$ when evaluating the impact of event ($a_E$,$i_E$,$T_E$,$l_E$).

\subsection{Extending the dynamical model with adaptation rules}\label{Extending the dynamical model with adaptation rules}
In the simulation, the adaptation of a parameter entry $v_{\text{orig}}$, which encompass as before $\alpha_{a,i}^p$, $\beta_{a,i}^p$, $\nu_{a,i}^p$, $T_{ab}^i$, $\eta_{a,i}^{\text{exp}}$ and $\eta_{a,i}^{\text{prod}}$, calculated based on the data for $2020$, the newest available year, by means of a weight adjustment $w$ and a rewiring $r$ after its associated sector $(a_S,i_s)$ has experienced a relative loss $l_S$ at $t=1$, one time step after the initialisation of the shock, of the simulation is applied as follows and results in $v_{\text{adap}}$, 
\begin{equation}
    v_{\text{adap}}=l_S \cdot w \cdot v_{\text{orig}} + l_S \cdot r \quad.
    \label{transformation_general}
\end{equation}
Using the example of the production output parameter $\alpha$ again, and using the adaptations derived in \eqref{alpha_adaptation_derivation} to \eqref{alpha_adaptation_aggregation}, the resulting parameter is
\begin{equation}
    \alpha_{a_S,i_S,\text{adap}}^p=l_S \cdot {W_\alpha}_{a_S,i_S}^p \cdot \alpha_{a_S,i_S,\text{orig}}^p + l_S \cdot {R_\alpha}_{a_S,i_S}^p \quad .
\end{equation}
$l_S$, the relative loss in available amount, must exceed the same limit $\Delta_{\text{rel}}$ as the events defined in \eqref{relative loss}. The absolute loss in available amount must also exceed the absolute event definition limit $\Delta_{\text{abs}}$ as in \eqref{absolute loss}. All nodes that fulfil these two criteria, denoted by 3-tuples ($a_S$,$i_S$,$l_S$), adapt their associated parameters at $t=1$. The parameters remain in the adapted state for the rest of the simulation, allowing the system to reach a stable state after a few time steps.\\

The adaptation rules fulfill a series of constraints. Entries in the trade matrix that would transform an item $i$ into item $j$ while being exported from country $a$ to country $b$ are always kept $0$, as are entries in the production output matrix that would denote an output item $i$ from a process $p$ in country $a$ to another country $b$.\\

The sum over all exports of an item originating from one country must equal one, so rows in $T_{ab, \text{adap}}^i$ are rescaled such that the ratio between the amounts that different importers $b$ receive and the the total exported amount from exporter $a$ remains constant,
\begin{equation}
    T_{ab, \text{adap,norm}}^i= \frac{T_{ab, \text{adap}}^i}{\displaystyle\sum_{(b,i) \in (\mathcal{A}\times\mathcal{I})}{T_{ab, \text{adap}}^i}} \quad.
\label{trade_adap_norm}  
\end{equation}
The production input parameter $\nu$ must fulfil a similar constraint, ensuring that not more than $100\%$ of a for production allocated available product is distributed towards production processes,
\begin{equation}
    \nu_{a,i,\text{adap,norm}}^p= \frac{\nu_{a,i,\text{adap}}^p}{\displaystyle\sum_{(a,p) \in (\mathcal{A}\times\mathcal{P})}{\nu_{a,i,\text{adap}}^p}} \quad.
\end{equation}
The shares that allocate the available amounts to different purposes cannot accumulate above or below one, so the shares associated with each purpose $k$ are scaled accordingly,
\begin{equation}
    \eta_{a,i,\text{adap,norm}}^k=\frac{\eta_{a,i,\text{adap}}^k}{\displaystyle\sum_k \eta_{a,i,\text{adap}}^k} \quad.
\end{equation}
The total output rate of a production process $p$ in country $a$ for all its products $i$ is kept constant at its original level,
\begin{equation}
    \alpha_{a,i,\text{adap,norm}}^p=\alpha_{a,i,\text{adap}}^p \frac{\displaystyle\sum_{(a,p) \in (\mathcal{A}\times\mathcal{P})}{\alpha_{a,i,\text{orig}}^p}}{\displaystyle\sum_{(a,p) \in (\mathcal{A}\times\mathcal{P})}{\alpha_{a,i,\text{adap}}^p}} \quad ,
\end{equation}
\begin{equation}
    \beta_{a,i,\text{adap,norm}}^p=\beta_{a,i,\text{adap}}^p \frac{\displaystyle\sum_{(a,p) \in (\mathcal{A}\times\mathcal{P})}{\beta_{a,i,\text{orig}}^p}}{\displaystyle\sum_{(a,p) \in (\mathcal{A}\times\mathcal{P})}{\beta_{a,i,\text{adap}}^p}} \quad .
\end{equation}

\subsection{Item substitutability}\label{Item substitutability}
Product substitutability is inferred from correlations between local losses in one product and import surges in another similar product. Products are considered similar, and therefore potential substitutes, if they belong to the same commodity group, see \ref{Commodity groups}, defined by [\hyperlink{19}{19}]. Given the set of relevant events ($a_E$,$i_E$,$T_E$,$l_E$) derived in \ref{Event definition}, we first define a general trade import index $I_{a_E}^j$ for each item $j$ similar to event item $i_E$ in event country $a_E$ by summing over all relevant import indices. To be included in the sum, sector $(a_E,i_E)$ must be a substantial importer from $(b,i_E)$ in question, receiving at least $0.1$ \% of the sector's exports,
\begin{equation}
    I_{a_E}^j (T_E)=\sum_b T_{a_Eb}^j (T_E) \quad .
\end{equation}
We derive the mean relative change in this trade-import index with respect to $j$ by averaging over all events affecting sectors of item $i_E$, giving the substitutability index $s_{i_Ej}$ as
\begin{equation}
        s_{i_Ej} = \biggl\langle \frac{I_{a_E}^j(T_E+1)-I_{a_E}^j(T_E)}{I_{a_E}^j(T_E)} \biggr\rangle_{a_E,j\neq i_E} \quad.
\end{equation}
The substitution rule is only considered in the model if the following null hypothesis $H_0^A$ is rejected,
\begin{equation}
H_0^A: \quad s_{i_Ej} \leq 0 \quad.
\end{equation}
In addition, a permutation test ensures the consistency of the results by randomly permuting over the distribution of event item-area combinations and correcting the resulting p-values using the Benjamini-Hochberg correction method [\hyperlink{38}{38}] to mitigate errors associated with multiple testing. The second null hypothesis $H_0^B$ tests whether the mean of the permutations of $s_{i_Ej}$ is less than 0,
\begin{equation}
H_0^B: \quad \langle s_{i_Ej} \rangle_{\text{perm}} \leq 0 \quad.
\end{equation}
For both null hypotheses, the significance level is set to $p<0.05$. As we here derive an item $\times$ item sized substitutability matrix $s$, to apply it to the trade parameter we inflate the size by applying the substitution scheme to every country $a$, resulting in the full substitutability index matrix $S$,
\begin{equation}
    s_{i_Ej} = S_{i_Ej}^a \quad \forall a \in \mathcal{A} \quad .
\end{equation}
The substitutability index adjustment is applied when a sector has lost a substantial proportion and quantity of its available goods and has not been able to recover through the adjustment applied at $t=1$. The substitutability indices are applied at $t=2$ in addition to the first round of adjustments. They are applied according to the same criteria as in \ref{Extending the dynamical model with adaptation rules}, which are derived from \eqref{relative loss} and \eqref{absolute loss}, to all entries that define the import parameters to the affected sector $(a_S,i_S)$ for which $j$ could serve as a substitute,
\begin{equation}
    T_{a_Sb,\text{adap}}^j = (1+S_{i_Sj}^a)\cdot T_{a_Sb,\text{orig}}^j \quad.
    \label{subsitute_application}
\end{equation}
$T_{a_Sb,\text{orig}}^j$ represent the current iteration of the trade matrix, as the result $T_{a_Sb,\text{adap,norm}}^j$ obtained after \eqref{trade_adap_norm} in the previous time-step. This method only adjusts existing trade relationships. This maintains the item-blocked structure of the trade matrix. However, \eqref{trade_adap_norm} must be reapplied to the result of \eqref{subsitute_application} to ensure that there are no forbidden gains or losses of goods within the trade.
\section{Quantifying the impact of adaptation}\label{Quantifying the impact of adaptation}
\subsection{Import trade adaptation}\label{Trade adaptation}
The estimated impact of each import trade adjustment multiplication rule ${M_T}_{ab}$ with country $a$ being the by a shock affected one is derived as follows by averaging over all items with existing trade flows from a country $a$ to $b$,
\begin{equation}
    \iota_{ab} = {M_T}_{ab} \cdot\langle T^i_{ab} \eta^{exp}_{a,i} {x_0}^i_a(0)\rangle_i \quad.
    \label{trade_weight_adjustment_impact}
\end{equation}
\subsection{Item substitution}\label{Item substitution}
When a loss of one item triggers the substitution step, the model strengthens all existing trade relationships for possible substitutes by the factor of the corresponding substitution index of the two products. Again, the trade share constraint will reduce the actual impact. The estimated impact of a substitution adjustment with $S_{ij}$ is given by
\begin{equation}
    \iota_{ij} = (1+S_{ij})\cdot\langle \sum_b T^j_{ab} \eta^{exp}_{a,j} x^j_a(0)\rangle_a \quad.
    \label{substitution_impact}
\end{equation}
\section{Cumulative losses across regions or product groups}\label{Cumulative losses across regions or product groups}
\eqref{absolute loss pc_shock} describe how the impact of a shock to sector $(a,i)$ is felt in sector $(b,j)$. Evaluating the severity of the same shock to an entire region $D$  concerning food item $j$ is done as follows,
\begin{equation}
    L_{a\rightarrow D}^{i\rightarrow j}(t)=\frac{\displaystyle\sum_{b \in D}\Bigl(\underline{x}_b^j(t)-x_{a\rightarrow b}^{i\rightarrow j}(t)\Bigr)}{\displaystyle\sum_{b \in D}z_b} \quad
    \label{loss_continent}
\end{equation}
If we want to evaluate the cumulative impact of a shock $(a,i)$ to every product within a group of food items $K$ in country $b$, the losses are calculated as follows,
\begin{equation}
    L_{a\rightarrow b}^{i\rightarrow K}(t)=\frac{\displaystyle\sum_{j \in K}\Bigl(\underline{x}_b^j(t)-x_{a\rightarrow b}^{i\rightarrow j}(t)\Bigr)}{z_b} \quad
    \label{loss_product_group}
\end{equation}
Combining both leads to,
\begin{equation}
    L_{a\rightarrow D}^{i\rightarrow K}(t)=\frac{\displaystyle\sum_{b \in D}\Bigl(\displaystyle\sum_{j \in K}\Bigl(\underline{x}_b^j(t)-x_{a\rightarrow b}^{i\rightarrow j}(t)\Bigr)\Bigr)}{\displaystyle\sum_{b \in D}z_b} \quad
    \label{loss_continent_product_group}
\end{equation}
\section{Evaluating effects of combined shocks}\label{Evaluating effects of combined shocks}
 We can compare the added losses of the individual shock scenario with the loss in the superposed scenario as follows. We first simulate two different scenarios:
 \begin{itemize}
    \item scenario $1$: A combined shock scenario with two simultaneous shocks resulting in $x_{a_1,a_2\rightarrow b}^{i_1,i_2\rightarrow j}$, representing the amount available in sector $(j,b)$ at the end of the simulation.
    \item scenario $2$: Two single shock scenarios, simulated independently, resulting in $x_{a_1\rightarrow b}^{i_1\rightarrow j} $ and $x_{a_2\rightarrow b}^{i_2\rightarrow j}$ which represent the amount available in sector $(j,b)$ at the end of each simulation.
    \end{itemize}
Then, we compare whether the two scenarios shock by subtracting from the losses in scenario $1$ the sum of losses in scenario $2$,
\begin{equation}
    SI_{a_1,a_2\rightarrow b}^{i_1,i_2\rightarrow j} = L_{a_1,a_2\rightarrow b}^{i_1,i_2\rightarrow j}-(L_{a_1\rightarrow b}^{i_1\rightarrow j} +L_{a_2\rightarrow b}^{i_2\rightarrow j}) \quad .
\end{equation}
If $SI_{a_1,a_2\rightarrow b}^{i_1,i_2\rightarrow j} > 0$, the losses indicate a super-additive shock effect. $SI_{a_1,a_2\rightarrow b}^{i_1,i_2\rightarrow j} < 0$ reveals a sub-additive impact. Being equal to $0$ indicates that there is no measurable effect of the superposition of two shocks.

To determine the general effect of superposing shocks on groups of products or countries, we insert \eqref{loss_continent_product_group} in \eqref{superposition_impact} and obtain the following formula for calculating the effect on group of countries  $D$ and group of products $K$,
\begin{equation}
    SI_{a_1,a_2\rightarrow D}^{i_1,i_2\rightarrow K} = L_{a_1,a_2\rightarrow D}^{i_1,i_2\rightarrow K}-(L_{a_1\rightarrow D}^{i_1\rightarrow K} +L_{a_2\rightarrow D}^{i_2\rightarrow K}) \quad .
    \label{superposition_impact_general}
\end{equation}
If we are interested in the superposition impact on the global availability of only the shocked products, $K=\{i_1,i_2\}$ and $D=\mathcal{A}$. To determine the total impact on all products and all countries, so the entire network, $K=\mathcal{I}$ and $D=\mathcal{A}$.\\

To get a general idea of how overlapping shocks, not the single combination of specific shocks $(a_1,i_1)$ and $(a_2,i_2)$, affect the losses to the shocked item and the network in general, we sample two random shocks $({a_r}_1,{i_r}_1)$ and $({a_r}_2,{i_r}_2)$ from a collection of possible major shocks to primary products and compare their aggregated results $SI_{{a_r}_1,{a_r}_2\rightarrow D}^{{i_r}_1,{i_r}_2\rightarrow K}$ for the shocked products ($K=\{{i_r}_1,{i_r}_2\}$ and $D=\mathcal{A}$) and the entire network ($K=\mathcal{I}$ and $D=\mathcal{A}$). Only shocks to sectors that pertain primary products and export are eligible. Among the shocks that meet these criteria, we consider the largest $100$ shocks according to their production volume. As a reference, Indian rice ranks $6$ and Ukrainian wheat ranks $56$ in the list according to this measure. Both products from the two shocks studied must always belong to the same product group in order for the superimposed substitution effect to take effect.
\newpage
\section*{Data Availability}
The data used in this study as input for simulations is available on GitHub,
\url{https://github.com/sokaba/adaptive_food_supply_network/tree/main/input}
\section*{Code Availability}
Python was used to perform the simulations and data analysis. Simulation and analysis code for this study is available in a repository at,
\url{https://github.com/sokaba/adaptive_food_supply_network}
\section*{Acknowledgements}
We are grateful to Giacomo Zelbi, Jan Fialkowski and Tobias Reisch for illuminating discussions. This work was supported by the Oesterreichische Nationalbank (OeNB) under anniversary fund project 18696, the Austrian Science Promotion Agency (FFG) under 873927, the Federal Ministry of the Republic of Austria for Climate Action, Environment, Energy, Mobility, Innovation and Technology (BMK) under GZ 2021-0.664.668 and the Federal Ministry of the Republic of Austria for Labour and Economy (BMAW) under Wi-2023-37526/3-E.
\section*{Author Contribution Statement}
Peter Klimek, Sophia Baum, Moritz Laber and Stefan Thurner contributed to study conception and design. Data was collected by Martin
Bruckner. Analysis was carried out by Sophia Baum. Results were interpreted by Peter Klimek, Sophia Baum and Stefan
Thurner. The artwork was conceived by Liuhuaying Yang, Sophia Baum and Moritz Laber. The first draft of the manuscript was written
by Sophia Baum and all authors commented on previous versions of the manuscript. All authors read and approved the final
manuscript.
\section*{Competing Interest Statement}
The authors declare no competing interests.
\newpage
\chapter{Bibliography}

\begin{enumerate}
\item \hypertarget{1}{\textit{WFP at a Glance - A guide to the facts, figures and frontline work of the World Food Programme}. 2024. \texttt{URL: }\url{https://www.wfp.org/publications/wfp-glance}.}
\item \hypertarget{2}{\textit{World Food Programme HungerMap}. 2024. \texttt{URL: }\url{https://hungermap.wfp.org/}.}
\item \hypertarget{3}{Benjamin Parkin et al. \textit{The return of the rice crisis}. 2023. \texttt{URL: }\url{https://www.ft.com/content/416736ec-7960-496d-b6c8-fd7a2fd99668}.}
\item \hypertarget{4}{\textit{Global Information and Early Warning System Country Briefs India.} 2023. \texttt{URL: }\url{https://www.fao.org/giews/countrybrief/country.jsp?code=IND}.}
\item \hypertarget{5}{\textit{United Nations Two-Year Update - Protection of civilians: impact of hostilities on civilians since 24 February 2022}. 2024. \texttt{URL: }\url{https://ukraine.un.org/en/261245-two-year-update-protection-civilians-impact-hostilities-civilians-24-february-2022}.}
\item \hypertarget{6}{\textit{Brief on the interruption of the Black Sea Grain Initiative and its potential implications on global food markets and food security}. 2023. \texttt{URL:} \url{https://openknowledge.fao.org/handle/20.500.14283/cc7271en}.}
\item \hypertarget{7}{Mária Ercsey-Ravasz et al. “Complexity of the international agro-food trade network and its impact on Food Safety”. In: \textit{PLOS ONE} 7.5 (2012), e37810. \texttt{DOI:} \url{https://doi.org/10.1371/journal.pone0037810}.}
\item \hypertarget{8}{Tiziano Distefano et al. “Shock transmission in the International Food Trade Network”. In: \textit{PLOS ONE} 13.8 (2018), pp. 1–15. \texttt{DOI: }\url{https://doi.org/10.1371/journal.pone.0200639}.}
\item \hypertarget{9}{Christopher Bren d’Amour et al. “Teleconnected food supply shocks”. In: \textit{Environmental Research Letters} 11.3 (2016), p. 035007. \texttt{DOI: }\url{https://doi.org/10.1088/1748-9326/11/3/035007}.}
\item \hypertarget{10}{Richard S. Cottrell et al. “Food production shocks across land and sea”. In: \textit{Nature News} (2019), pp. 130–137. \texttt{DOI: }\url{https://doi.org/10.1038/s41893-018-0210-1}.}
\item \hypertarget{11}{Jacob Schewe, Christian Otto, and Katja Frieler. “The role of storage dynamics in annual wheat prices”. In: \textit{Environmental Research Letters} 12.5 (2017), p. 054005. \texttt{DOI: }\url{https://doi.org/10.1088/1748-9326/aa678e}.}
\item \hypertarget{12}{Derek Headey. “Rethinking the global food crisis: The role of trade shocks”. In: \textit{Food Policy} 36.2 (2011), pp. 136–146. \texttt{DOI: }\url{https://doi.org/10.1016/j.foodpol.2010.10.003}.}
\item \hypertarget{13}{Marco Lagi et al. “The Food Crises: A quantitative model of food prices including speculators and ethanol conversion”. In: \textit{SSRN Electronic Journal} (2011). \texttt{DOI:} \url{https://doi.org/10.2139/ssrn.1932247}.}
\item \hypertarget{14}{Getaw Tadesse et al. “Drivers and triggers of international food price spikes and volatility”. In: \textit{Food Policy} 47 (2014), pp. 117–128. \texttt{DOI:} \url{https://doi.org/10.1016/j.foodpol.2010.10.003}.}
\item \hypertarget{15}{Jessica A Gephart et al. “Vulnerability to shocks in the global seafood trade network”. In: \textit{Environmental Research Letters} 11.3 (2016), p. 035008. \texttt{DOI:} \url{https://doi.org/10.1088/1748-9326/11/3/035008}.}
\item \hypertarget{16}{Rebekka Burkholz and Frank Schweitzer. “International crop trade networks: the impact of shocks and cascades”. In: \textit{Environmental Research Letters} 14.11 (2019), p. 114013. \texttt{DOI:} \url{https://doi.org/10.1088/1748-9326/ab4864}.}
\item \hypertarget{17}{Marie-Cécile Dupas, José Halloy, and Petros Chatzimpiros. “Time dynamics and invariant subnetwork structures in the world cereals trade network”. In: \textit{PLOS ONE} 14.5 (2019), pp. 1–21. \texttt{DOI:} \url{https://doi.org/10.1371/journal.pone.0216318}.}
\item \hypertarget{18}{Kathyrn R. Fair, Chris T. Bauch, and Madhur Anand. “Dynamics of the Global Wheat Trade Network and Resilience to Shocks”. In: \textit{Scientific Reports} 7.1 (2017), p. 7177. \texttt{DOI:} \url{https://doi.org/10.1038/s41598-017-07202-y}.}
\item \hypertarget{19}{Martin Bruckner et al. “Fabio—the construction of the Food and agriculture biomass input–output model”. In: \textit{Environ. Sci. Technol.} 53.19 (2019), pp. 11302–11312. \texttt{DOI:} \url{https://doi.org/10.1021/acs.est.9b03554}.}
\item \hypertarget{20}{Moritz Laber et al. “Shock propagation from the Russia–Ukraine conflict on International Multilayer Food Production Network Determines Global Food Availability”. In: \textit{Nature Food} 4.6 (2023), pp. 508–517. \texttt{DOI:} \url{https://doi.org/10.1038/s43016-023-00771-4}.}
\item \hypertarget{21}{Philippe Marchand et al. “Reserves and trade jointly determine exposure to food supply shocks”. In: \textit{Environmental Research Letters} 11.9 (2016), p. 095009. \texttt{DOI:} \url{https://doi.org/10.1088/1748-9326/11/9/095009}.}
\item \hypertarget{22}{Ronald E. Miller and Peter D. Blair. \textit{Input-Output Analysis: Foundations and Extensions}. 2nd ed. Cambridge University Press, 2009}
\item \hypertarget{23}{ Luca Galbusera and Georgios Giannopoulos. “On input-output economic models in disaster impact assessment”. In: \textit{International Journal of Disaster Risk Reduction} 30 (2018), pp. 186–198. \texttt{DOI:} \url{https://doi.org/10.1016/j.ijdrr.2018.04.030}.}
\item \hypertarget{24}{\textit{The State of Agricultural Commodity Markets 2022. The geography of food and agricultural trade:Policy approaches for sustainable development}. 2022. \texttt{DOI:} \url{https://doi.org/10.4060/cc0471en}.}
\item \hypertarget{25}{Michael J Puma et al. “Assessing the evolving fragility of the global food system”. In: \textit{Environmental Research Letters} 10.2 (2015), p. 024007. \texttt{DOI:} \url{https://doi.org/10.1088/1748-9326/10/2/024007}.}
\item \hypertarget{26}{Anuradha Mittal. “The 2008 Food Price Crisis: Rethinking Food Security Policies”. In: \textit{G-24 Discussion Paper Series} 56 (2009).}
\item \hypertarget{27}{Kyle Frankel Davis, Shauna Downs, and Jessica A. Gephart. “Towards Food Supply Chain Resilience to environmental shocks”. In: \textit{Nature Food} 2.1 (2020), pp. 54–65. \texttt{DOI:} \url{https://doi.org/10.1038/s43016-020-00196-3}.}
\item \hypertarget{28}{Leonie Wenz and Anders Levermann. “Enhanced economic connectivity to foster heat stress–related losses”. In: \textit{Science Advances} 2.6 (2016), e1501026. \texttt{DOI:} \url{https://doi.org/10.1126/sciadv.1501026}.}
\item \hypertarget{29}{\textit{United Nations, Department of Economic and Social Affairs, Population Division - World Population Prospects 2022} (2022). \texttt{URL:} \url{ https://population.un.org/wpp/Download/Standard/MostUsed/}.}
\item \hypertarget{30}{\textit{United Nations Development Programme - Human Development Report 2021-22}. (2022). \texttt{URL:} \url{http://report.hdr.undp.org}.}
\item \hypertarget{31}{Jesus Felipe et al. “Product complexity and economic development”. In: \textit{Structural Change and Economic Dynamics} 23.1 (2012), pp. 36–68. \texttt{DOI:} \url{https://doi.org/10.1016/j.strueco.2011.08.003}.}
\item \hypertarget{32}{Samira Choudhury and Derek Headey. “What drives diversification of national food supplies? A cross-country analysis”. In: \textit{Global Food Security} 15 (2017), pp. 85–93. \texttt{DOI:} \url{https://doi.org/10.1016/j.gfs.2017.05.005}.}
\item \hypertarget{33}{Jukka Korpela et al. “An analytic approach to production capacity allocation and supply chain design”. In: \textit{International Journal of Production Economics} 78.2 (2002), pp. 187–195. \texttt{DOI:} \url{https://doi.org/10.1016/S0925-5273(01)00101-3}.}
\item \hypertarget{34}{Christian Diem et al. “Quantifying firm-level economic systemic risk from nation-wide supply networks”. In: \textit{Scientific Reports} 12.1 (2022), p. 7719. \texttt{DOI:} \url{https://doi.org/10.1038/s41598-022-11522-z}.}
\item \hypertarget{35}{Michael Obersteiner et al. “The phosphorus trilemma”. In: \textit{Nature Geoscience} 12.1 (2013),pp. 897–898. \texttt{DOI:} \url{https://doi.org/10.1038/ngeo1990}.}
\item \hypertarget{36}{Richard Oloruntoba, Ramaswami Sridharan, and Graydon Davison. “A proposed framework of key activities and processes in the preparedness and recovery phases of disaster management”. In: \textit{Disasters} 42.3 (2018), pp. 541–570. \texttt{DOI:} \url{https://doi.org/10.1111/disa.12268}.}
\item \hypertarget{37}{Asjad Naqvi, Franziska Gaupp, and Stefan Hochrainer-Stigler. “The risk and consequences of multiple breadbasket failures: an integrated copula and multilayer agent-based modeling approach”. In: OR Spectrum 42.3 (Sept. 2020), pp. 727–754. \texttt{DOI:} \url{https://doi.org/10.1007/s00291-020-00574-0}.}
\item \hypertarget{38}{Yoav Benjamini and Yosef Hochberg. “Controlling the False Discovery Rate: A Practical and Powerful Approach to Multiple Testing”. In: \textit{Journal of the Royal Statistical Society: Series B (Methodological)} 57.1 (1995), pp. 289–300. \texttt{DOI:} \url{https://doi.org/10.1111/j.2517-6161.1995.tb02031.x}.}
\item \hypertarget{39}{Brain W. Matthews. “Comparison of the predicted and observed secondary structure of T4 phage lysozyme”. In: \textit{Biochimica et Biophysica Acta (BBA) - Protein Structure} 405.2 (1975), pp. 442–451. \texttt{DOI:} \url{https://doi.org/10.1016/0005-2795(75)90109-9}.}
\end{enumerate}
\newpage
\chapter{Supplementary information}
\section{Multilayer network model of global food production and trade}\label{Multilayer network model of global food production and trade}
\begin{figure}[h!]
\vspace{1cm}
\centering
    \includegraphics[width=\linewidth, valign=t]{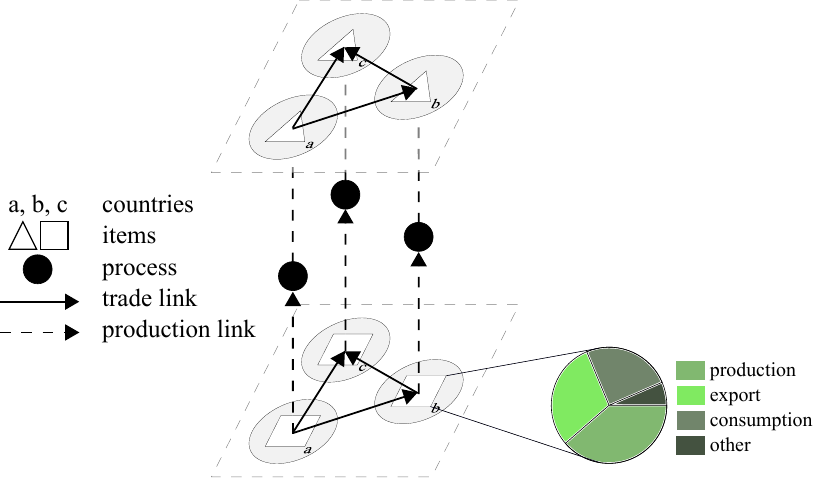}
\caption{Schematic representation of the shock propagation model as a multilayer bipartite network of global food production and trade. We show an example for three countries and two products. Each time step of the simulation goes through three iterative steps: trade, production and allocation. A layer represents a food product that can only be traded within the layer (trade links as solid arrows). Transitions between layers occur only through a production process (production process links as dotted lines). For each country-product combination, also called sector, fractions are individually allocated to different purposes (shown as a pie chart for an exemplary country-product combination). Note that each process may have multiple inputs and outputs, which are not explicitly shown here.}
\label{model}
\end{figure}

\begin{figure}[t!]
\vspace{5cm}
    \centering
    \includegraphics[width=\linewidth]{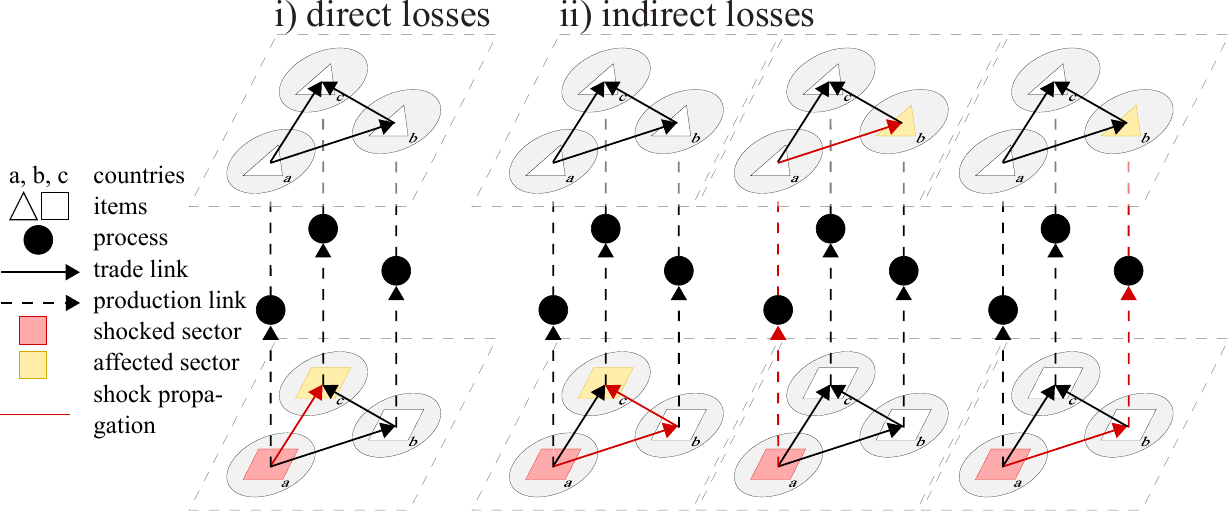}
    \caption{Shock propagation through the model allows for four types of shock transmission (i). Direct losses at $t=1$ occur immediately after the shock to production at $t=0$, which can only affect direct trading partners (ii). These direct effects can propagate within a layer of the trade network and lead to indirect trade losses in third countries (first figure). Other items may also suffer losses, either because production in the shocked area and hence trade in the secondary item is affected (second figure), or because production in the receiving areas is affected (third figure).}
    \label{shock_propagation}
\vspace{6cm}
\end{figure}

The adjustment rules are based on the parameters of a multilayer bipartite network model of global food trade and production described by Laber et al. [\hyperlink{20}{20}], calculated for several years. Figure \ref{model} and \ref{shock_propagation} show a schematic abstract sketch of the model, with index and shapes representing countries and items as opposed to Figure \ref{model overview} with describes a specific scenario.
\newpage
\section{Stability of application rules and optimal event limits}\label{Stability of application rules and optimal event limits}
If the adjustment rules prove to be stable over time, it is reasonable to assume that they can also be used to estimate future compensation. To test their stability over time, the years for which the data set is available are divided into two parts. The first part spans from $1992$ to $2006$ and the second part from $2006$ to $2020$. The derivation of the adjustment rules, including \ref{Event definition} and \ref{Calibration of adaptation rules}, is performed for event years at the beginning and end of each section, which are reduced from five to three years. This adjustment ensures that each section has eight years instead of four as potential event times for comparison. The resulting correlation coefficients between the two resulting transformation matrices for each parameter (both multiplier and rewiring adjustment) are calculated for different sets of event definition limits and then compared to obtain the optimal limit. The final event limits should also take into account that there are enough relevant events within the limits to cover all $192$ countries and $123$ products included in the model, while ensuring that the computation time remains reasonable.\\

To find the optimal limit, we first analyse the general trends when varying the limits for the three event criteria. Keeping the absolute limit as low as possible results in the highest number of events and also shows the steepest decline in events. We choose $1000$ tonnes as the absolute limit because it is rather low compared to the largest observed amounts available in the model, which can reach up to $1.5\times10^{10}$ tonnes, but will exclude some minor fluctuations that account for only a few tonnes. The distributions for the relative and deviation limits show the steepest decreases and increases around $0.26$ and $0.32$ respectively. For these limits, the number of events is also sufficiently high over the whole and the two separate periods of observation. Keeping one of the bounds in question fixed, while varying the other from $0.1$ to $0.7$, the fitting matrices for each parameter are calculated for both parts of the separated time period. For each pair of identical bounds and parameters, the Matthews correlation [\hyperlink{39}{39}], a measure of binary association, is determined. This type of correlation measure does not take into account the size of an entry in the transition matrix; rather, it assesses whether an entry exists by measuring the correlation of whether a rule itself is present in both, neither, or only one section. The matrices derived by the method introduced in \ref{Calibration of adaptation rules} show a high to very high correlation for every kind of combination of boundary definitions. Only the substitutability index \ref{Item substitutability} is not highly correlated over the whole range of limits and only reaches moderate correlation for relative drop limits below $0.3$ and allowed deviation limits above $0.3$, corresponding to our initially chosen values of $\Delta_{\text{rel}} = 0.26$ and $\Delta_{\text{dev}}= 0.32$.
\begin{figure}[H]
    \vspace{2cm}
    \includegraphics[width=\linewidth]{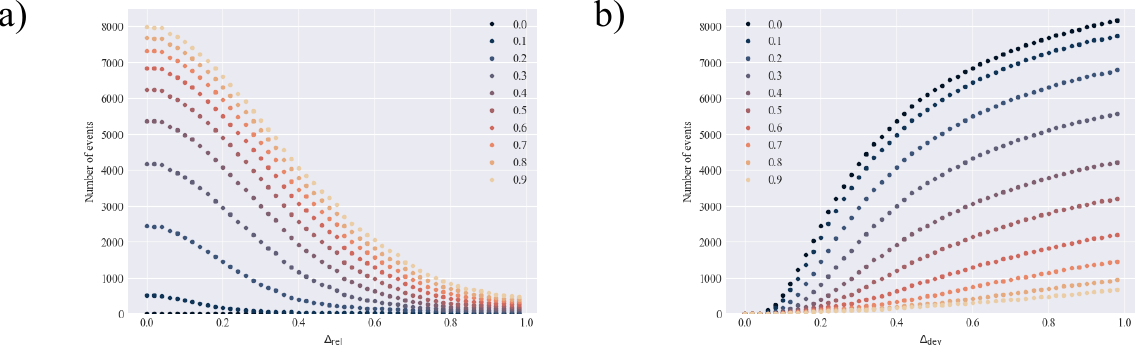}
\caption{Number of events for different sets of relative drop and deviation limits (a). The relative limit is varied from $0$ to $1$ for different levels of allowed deviation from $0.0$ to $0.9$ (b). The deviation limit is varied from $0$ to $1$ for different levels of the relative drop limit from $0.0$ to $0.9$. In general, the lower the relative drop limit and the higher the allowed deviation limit, the more relevant events can be observed. There is no clear transition point where the shape of the distribution changes, but the average steepest decline is observed for a relative drop limit of $0.26$ and the average steepest rise is observed for an allowed deviation limit of $0.32$.}
  \vspace{2cm}
    \includegraphics[width=\linewidth]{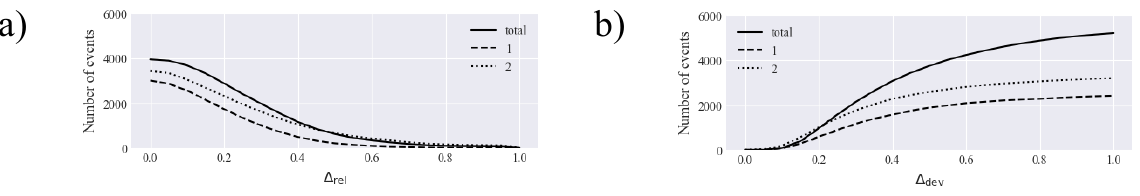}
\caption{Number of events within different time frames for different sets of relative drop and deviation limits: On the left, the allowable deviation limit is set to $0.32$ and the relative limit is varied from $0$ to $1$. On the left, the relative drop limit is set to $0.26$ and the deviation limit is varied from $0$ to $1$. Both distributions show that the combination of a deviation limit of $0.32$ and a relative limit of $0.26$ provides a high enough number of events for the total and both parts of the split period with counts well above $1000$.}
\vspace{2cm}
\end{figure}
\begin{figure}[H]
    \vspace{4cm}
    \includegraphics[width=\linewidth]{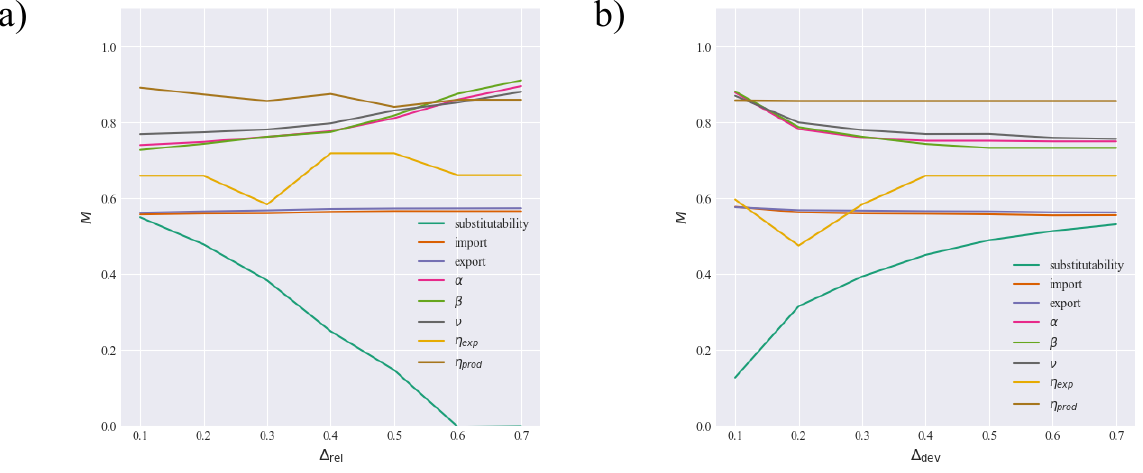}
\caption{Matthews correlation of the adaptation matrices of both time sections for varying (a) relative limit and (b) deviation limit. The substitution index (\ref{Item substitutability}) is the least correlated adaptation parameter over both bounds. When varying the relative limit (while keeping $\Delta_{\text{dev}} = 0.32$), the Matthews correlation for the substitution index falls below $0.4$, which we consider to be the threshold for moderate correlation, between $0.2$ and $0.3$, while all other adaptation matrices show strong to very strong correlation throughout. For varying deviation (keeping $\Delta_{\text{rel}} = 0.26$), the substitution index again shows the weakest correlation, exceeding $0.4$ around $0.3$. The other adaptation matrices again show moderate to high correlations throughout.}
\vspace{4cm}
\end{figure}
\newpage
\section{Data reconciliation}\label{Data Reconciliation}
\begin{figure}[!htb]
\vspace{2cm}
    \includegraphics[width=0.9\linewidth, valign=t]{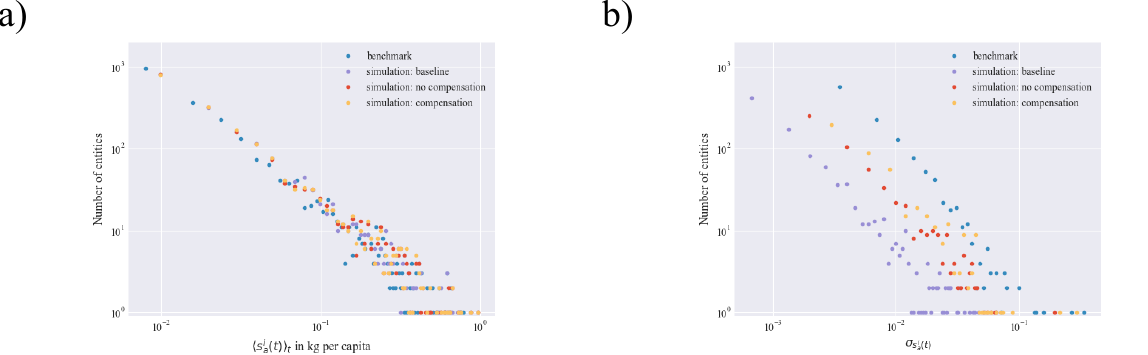}
\caption{Comparison of a country's (a) mean market share $\langle s_a^i(t) \rangle_t$ of an item over the duration of the simulation and (b) $\sigma_{s_a^i(t)}$. The mean shows similar distributions for all three simulations (violett: baseline, red: no compensation or static, yellow: compensation or adaptive) included in the graph, which are again very similar to the benchmark (blue). As for the standard deviation, the values for the benchmark scenario are much higher than in the baseline or shock simulation. The compensation scenario shows the highest standard deviations among the simulation scenarios and is therefore closest to the benchmark.}
\vspace{1cm}
\label{validation - market share}
    \includegraphics[width=0.9\linewidth, valign=t]{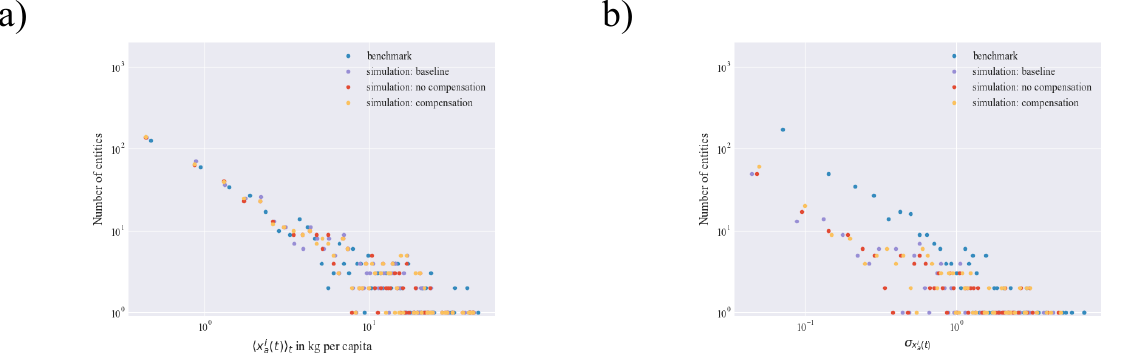}
\caption{Comparison of the (a) mean available volume per capita $\langle x_a^i(t) \rangle_t$ in a sector $(a,i)$ over the duration of the simulation and (b) its standard deviation $\sigma_{x_a^i(t)}$. The mean again shows similar distributions for all three simulations (violett: baseline, red: no compensation or static, yellow: compensation or adaptive) included in the graph, which are once more very similar to the benchmark (blue). As for the standard deviation, the values for the simulations follow similar distributions, each of them clearly with values smaller than the benchmark and not properly distinct from one another.}
\label{validation - pc}
\vspace{2.5cm}
\end{figure}

To validate the simulation results, comparing them with real-world data would be ideal. Unfortunately, we lack specific data describing the effects of singular shock propagation or their compensation mechanisms, apart from our base dataset, FABIO [\hyperlink{19}{19}]. We therefore make use of this available dataset and the resulting parameters, which include a starting vector ${x_0}_a^i$ for each year. Instead of deriving parameters from the usual $29$ years ($1992$ to $2020$), we base the adjustment rules on the first $19$ years ($1992$ to $2010$) and use the remaining years ($2011$ to $2020$) as a benchmark. The benchmark vectors, consisting of the initial vectors ${x_0}_a^i$, are normalised with respect to the first year included in the benchmark, $2011$, similar to equation \ref{x_0_norm} but using $2011$ instead of $1992$.\\
\\
For the simulation scenarios, we calculate the baseline scenario as usual, using the model parameters derived from the data for $2011$. For the two shocked scenarios (no compensation and compensation), the initiated shock consists of the real events of $2011$ observed in the data. Each event in $2011$ that fulfils the criteria of a relative loss greater than $0.26$, an absolute loss greater than $1000$ tonnes, and a maximum deviation of $0.32$ before and after the event is considered. According to the obtained set of shock events $(a,i,l)$ for the affected item $i$ in area $a$ with a relative drop of $l$, the reduced production outputs are derived using equation \ref{shock initiation} where $l$ replaces $\phi$. Production is reduced at each time step $t$ of the simulation,
\begin{equation}
    o_a^i (t)=(1-l_a^i) \cdot o_a^i (t) \quad \forall t \in \{0, ... , \tau\} \quad .
\end{equation}
The static scenario is run as described in section \ref{Model without compensation} using the model parameters derived from the $2011$ data and the events described above. In addition, the adaptive scenario applies the adjustment rules described in sections \ref{Extending the dynamical model with adaptation rules} and \ref{Item substitutability}. The results are shown in figures \ref{validation - market share} and \ref{validation - pc}. The market share of a sector $(a,i)$ is calculated as follows,
\begin{equation}
    s_a^i(t)=\frac{x_a^i(t)}{\displaystyle\sum_{a\in\mathcal{A}}x_a^i(t)} \quad .
\end{equation}Comparing the average market share $\langle s_a^i(t) \rangle_t$ of a country for an item over the simulation period and its standard deviation $\sigma_{s_a^i(t)}$, it can be seen that the scenario allowing for adjustment models the fluctuations in the available quantity more realistically than the static model. However, the trend in the standard deviation $\sigma_{x_a^i(t)}$ of the per capita volume of an item in a region over the simulation period is inconclusive. The baseline and the two shock scenarios show substantially lower dispersion than the benchmark scenario, but they cannot be properly distinguished from each other. Both $\langle s_a^i(t) \rangle_t$  and $\langle x_a^i(t) \rangle_t$  time series roughly follow the distribution of the benchmark scenario.
\newpage
\section{Adaptation rules in details}\label{Adaptation rules in details}
\subsection{Adaptation rule coverage}\label{Adaptation rule coverage}
We consider two classes of adaptation rules, multiplying (changing the weight of network links) or rewiring (creating new links). Comparing the number of rules discovered, multiplication is much more common than rewiring (Table \ref{adaptation_rule_table}), although each of the matrix pairs is based on the same set of parameter matrices and could therefore in principle have the same number of adaptation rules. We adapt every parameter, when triggering the adaptation, but only the adaptation of a select few will actually improve the food availability situation in the adapting sector, most notably the import direction of trade as further discussed in Section \ref{Adaptation Strategies}.
\begin{table}[h]
    \centering
    \caption{Number of rules for each adaptation method. For each parameter, except $\eta_{exp}$, the number of possible multiplication adaptations is at least one order of magnitude greater than the number of rewiring adaptations. We observe the largest number of possible applicable rules for both directions of trade parameters.} 
    \begin{tabular}{ | c | c | c | } 
         \hline
         \textbf{Variable} & \textbf{Number of multipliers} & \textbf{Number of rewirers}\\ 
         \hline
         \hline
         $T$ (import) & $2.786.934$ & $803.190$ \\ 
         $T$ (export) & $2.914.484$ & $876.744$ \\ 
         $\alpha$ & $22.080$ & $2.304$\\
         $\beta$ & $21.120$ & $384$\\
         $\nu$ & $169.728$ & $29.568$\\
         $\eta^{\text{exp}}$ & $21.696$ & $10.368$\\
         $\eta^{\text{prod}}$ & $14.208$ & $2.496$\\
         \hline
    \end{tabular}
    \label{adaptation_rule_table}
\end{table}
\newpage
\subsection{Trade adaptation}
\begin{figure}[!htb]
    \centering
    \includegraphics[width=0.8\linewidth]{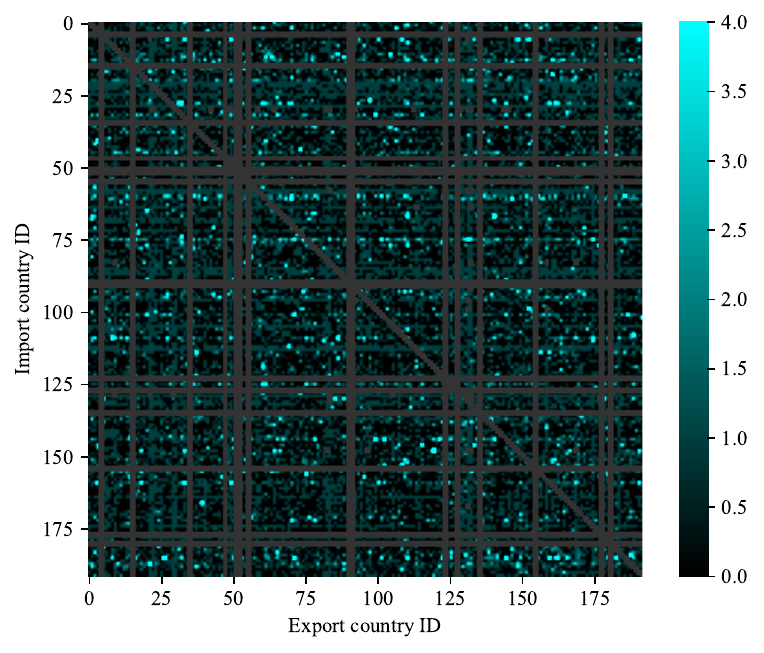}
    \label{trade_import_figure}
    \caption{Import multiplier matrix $W_{T_{ab}}$. A value of $1$, indicated by a dark cyan colour on the spectrum, represents no change in the multiplier for that parameter when the adjustment is triggered. Anything lighter indicates in increase when triggered. Dark grey areas indicate that this trade link between the two countries does not exist for any type of product. This mostly includes former nations that are still mentioned in the model due to their past existence between $1986$ and $2020$, but we have decided not to derive rules for them as an incomplete timeline would distort the event definition principles.}
\end{figure}

We only highlight the import direction adjustment in more detail in the paper (see Section section \ref{Adaptation Strategies}). The export direction is not as interesting as the import direction due to the structure of the model. Once a country has allocated a certain amount of an item for export, it is irretrievably removed from the available amount and exported. Changing this allocation requires changing the allocation shares of $\eta^{exp}$, not the trade matrix itself. From our simplistic perspective, a country only cares about how much it exports, not how those exports are distributed around the world, as this does not increase the amount of product available within the country.\\
The results for weight import adjustments, the import weight adjustment matrix, ${W_T}_{ab}$, are summarised in Fig. \ref{trade_import_figure}, which shows the aggregated country-to-country adjustment rules averaged over all food products. When importer $a$ faces a shortage of any given product, we find that $a$ scales its trade relations with trading partner $b$ in that product according to the entry in the matrix.
\newpage
\subsection{Production adaptation}
\begin{figure}[!htb]
 \includegraphics[width=\linewidth]{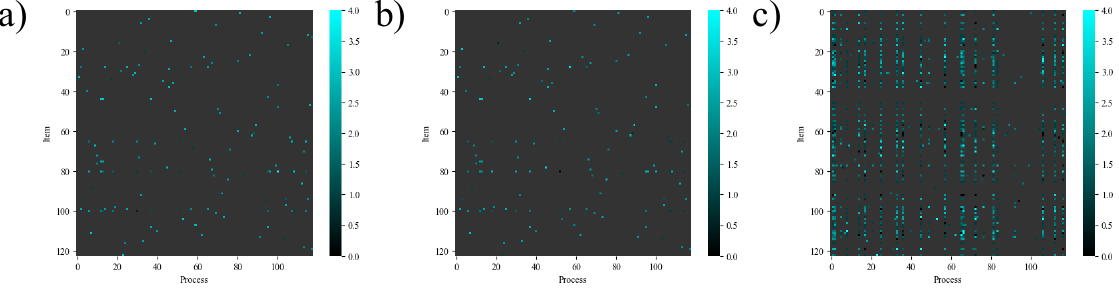}
\caption{Production multiplier matrices (a) ${M_\alpha}^k_i$ for production with input output rates, (b) ${M_\beta}^k_i$ for production without input output rates and  (c) ${M_\nu}^k_i$ for production without input output rates. A value of $1$, indicated by a cool blue-violet colour on the spectrum, represents no change by the multiplier for that parameter when the adjustment is triggered. Dark grey areas indicate that this production link between a production process and a product does not exist in any country. In particular, the output parameters $\alpha$ and $\beta$ and their adjustment matrices are rather sparse. On the input side, a handful of production processes accept a wide range of inputs, while others are limited to one possible input.}
\label{production_figure}
\end{figure}

Adjusting the distribution of production inputs and outputs more favourably may be another fruitful strategy for dealing with food shortages. Both the distribution of available items to and from production processes can affect the final quantities, as described in \ref{Trade and production adaptation strategies}. Figure \ref{production_figure} shows the aggregated item-process adjustment for each of the three production parameters $\alpha$, $\beta$ and $\nu$, averaged over all countries. $\alpha$ and $\beta$ describe the production rate processes, distinguishing only whether a process requires input items or not, while $\nu$ shows the proportions of how countries allocate items for production in general to production processes.\\

As in the case of trade adjustment, we have to take into account similar aspects that reduce the actual effects of adjustment. The base value to which we apply the multiplier may seem rather low, and in both cases of changing the distribution of inputs and outputs, we have to take into account the constraints of limited available inputs and maximum capacity of the production process. To estimate the maximum potential of a production adjustment, we derive the estimated impact of a production adjustment multiplication rule ${M}^k_i$ as follows, by averaging over all countries with existing production inflows or outflows between a process $k$ and an item $i$,
\vbox{
\baselineskip=20pt
\begin{align}
    {\iota_{\alpha}}^k_i &= {M_\alpha}^k_i \cdot\langle \alpha^k_{a,i} 
    \sum_j\nu^k_{a,j}\eta^{prod}_{a,j} x^j_a(0)\rangle_a \quad ,\\
    {\iota_{\beta}}^k_i &= {M_\beta}^k_i \cdot\langle \beta^k_{a,i} \rangle_a \quad , \\
    {\iota_{\nu}}^k_i &= {M_\nu}^k_i \cdot\langle \nu^k_{a,i}\eta^{prod}_{a,i} x^i_a(0)\rangle_a \quad .
\end{align}
}
Table \ref{production_table} shows the results. Among the top 3 most effective adaptation rules, all multipliers are in the interval $[1.78, 2.86]$, covering a much smaller range compared to trade adaptation. For $\alpha$ and $\beta$, the top 3 include the same item-process combinations, with \textit{Poultry Birds - Poultry Birds Farming} topping the list by far. In general, processes involving live animals seem to be favoured by the adaptation mechanism.
\newpage
\begin{table}[!htb]
\centering
\caption{Top $3$ impactful multiplier for each production parameter for (a) ${M_\alpha}^k_i$ representing the multiplier applied to production output with input links from process $k$ to item $i$ whenever any country experiences a substantial shortage of product $a$ according to the estimated impact ${\iota_{\alpha}}^k_i$, (b) ${M_\beta}^k_i$, which is the multiplier to be applied to output without input linkages from process $k$ to item $i$ whenever any country experiences a substantial shortage of product $a$ according to the estimated impact ${\iota_{\beta}}^k_i$; and (c) ${M_\nu}^k_i$ which is the multiplier to be applied to production input links from item $i$ to process $k$ whenever any country experiences a substantial shortage of product $a$ according to the estimated impact ${\iota_{\nu}}^k_i$. \\
* Live animals are measured in 1000 heads, not tonnes. Therefore, poultry is heavily favoured in the calculation of estimated impact because it is smaller than other farm animals such as pigs or cows and therefore occurs in larger numbers.}
\begin{tabular}{ p{0.25cm} |  p{4cm} |  p{4cm} |  C{1cm} |  C{4cm} | } 
 \cline{2-5}
 \multirow{2}{*}{\textbf{a)}}&\multirow{2}{*}{\textbf{Output item}} & \multirow{2}{*}{\textbf{Production process}} & \multirow{2}{*}{\boldmath{${M_\alpha}^k_i$}} & \textbf{Estimated impact \boldmath{${\iota_{\alpha}}^k_i$}} \\
  & & & & \textbf{in \boldmath{$10^7$} tons} \\
\cline{2-5}
 & Poultry Birds* & Poultry Birds farming & $2.86$ & $118.04$\\ 
 & Sugar cane & Sugar cane production & $2.85$& $2.77$\\ 
 & Pigs* & Pigs farming & $2.84$& $1.89$\\ 
 \cline{2-5}
 \end{tabular}
\begin{tabular}{ p{0.25cm} |  p{4cm} |  p{4cm} |  C{1cm} |  C{4cm} | } 
 \cline{2-5}
 \multirow{2}{*}{\textbf{b)}}& \multirow{2}{*}{\textbf{Output item}} & \multirow{2}{*}{\textbf{Production process}} & \multirow{2}{*}{\boldmath{${M_\beta}^k_i$}} & \textbf{Estimated impact \boldmath{${\iota_{\beta}}^k_i$}} \\
  & & & & \textbf{in \boldmath{$10^7$} tons} \\
 \cline{2-5}
 & Poultry Birds* & Poultry Birds farming & $2.50$ & $103.11$\\ 
 & Sugar cane & Sugar cane production & $2.39$& $2.33$\\ 
 & Pigs* & Pigs farming & $2.76$& $1.83$\\ 
 \cline{2-5}
 \end{tabular}
\begin{tabular}{ p{0.25cm} |  p{4cm} |  p{4cm} |  C{1cm} |  C{4cm} | } 
 \cline{2-5}
 \multirow{2}{*}{\textbf{c)}}& \multirow{2}{*}{\textbf{Input item}} & \multirow{2}{*}{\textbf{Production process}} & \multirow{2}{*}{\boldmath{${M_\nu}^k_i$}} & \textbf{Estimated impact \boldmath{${\iota_{\beta}}^k_i$}} \\
  & & & & \textbf{in \boldmath{$10^8$} tons} \\
 \cline{2-5}
 & Poultry Birds* & Poultry slaughtering & $1.78$ & $7.31$\\ 
 & Grazing & Dairy sheep husbandry & $2.55$ & $2.21$\\ 
 & Grazing & camels husbandry & $2.44$ & $2.11$\\ 
 \cline{2-5}
\end{tabular}
\label{production_table}
\end{table}
\newpage
\section{Extended results of isolated shock scenarios: India and Ukraine}\label{Extended results of isolated shock scenarios: India and Ukraine}

We demonstrated the model by considering two isolated shock scenarios: a complete shock to Indian rice exports and Ukrainian wheat exports. For both scenarios, we analyze two conditions: one where the affected sectors can adapt by adjusting trade and production parameters (\textit{India\_adap} and \textit{Ukraine\_adap}), and one where all parameters remain static throughout the simulation (\textit{India\_stat} and \textit{Ukraine\_stat}).\\

\begin{figure}[t]
\centering
  \includegraphics[width=\linewidth]{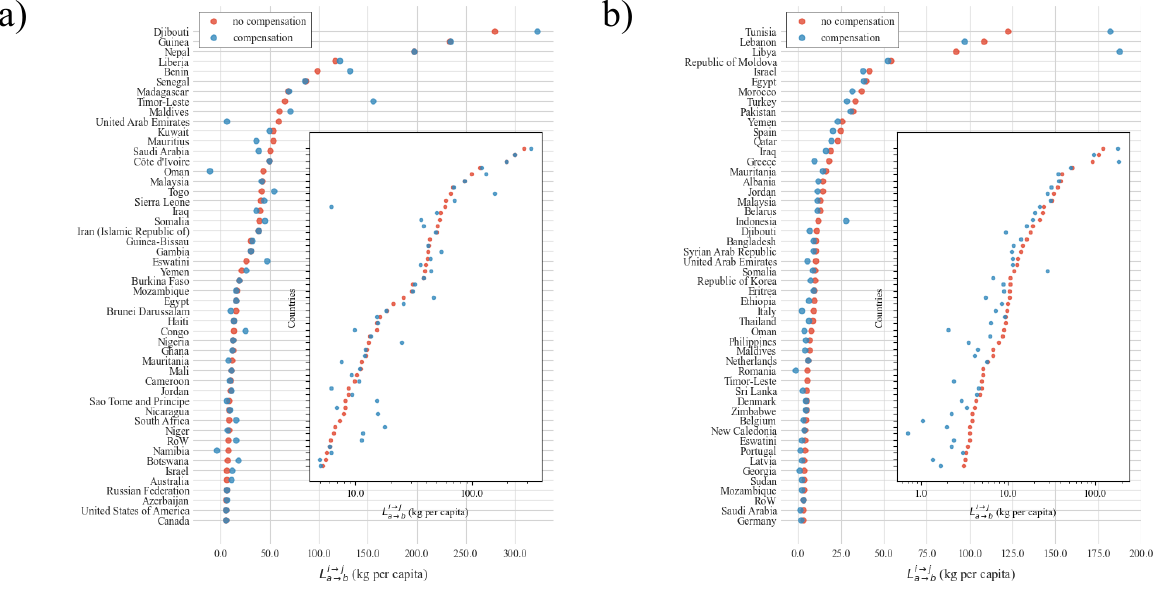}
\caption{ $L_{a\rightarrow b}^{i\rightarrow j}$ in the static (red) and adaptive (blue) scenario for the top $50$ losers in the static scenario: a) Shock to Indian rice ($a=\text{\texttt{IND}}$, $i,j=\text{\texttt{Rice and products}}$): Some of the worst affected countries can compensate for some of their losses, such as the United Arab Emirates, Kuwait, Mauritius, Saudi Arabia, or Oman. However, many others, such as Djibouti, Liberia, Benin, Timor-Leste, or the Maldives are performing worse. On the linear scale, it seems as the less affected nodes show barely any effect, but the logarithmic scale (inset) reveals that the lower nodes are relatively similarly affected. b) Shock to Ukrainian wheat ($a=\text{\texttt{UKR}}$, $i,j=\text{\texttt{Wheat and products}}$): The most affected countries, except for Tunisia, Libya, or Indonesia, are able to reduce their losses through adapted strategies. Again, the compensation effect seems rather limited on the linear scale for less affected nodes, but the logarithmic scale (inset) reveals the relatively large gains on the lower end.}
\label{top50}
\end{figure}

\textbf{Indian Rice Scenario:} In the simulation scenario ($a=\text{\texttt{IND}}$, $i=\text{\texttt{Rice and products}}$), we shock $100$\% of the Indian rice production, resulting in India's rice exports being diminished. The evaluation will focus solely on investigating losses in directly or indirectly affected countries other than India. Figure \ref{top50} a) illustrates the uncompensated losses $L_{a\rightarrow b}^{i\rightarrow j}$ for ($a=\text{\texttt{IND}}$, $i,j=\text{\texttt{Rice and products}}$) for the $50$ most affected countries in (\textit{India\_stat} and the impact of the adaptations, corresponding to Figure \ref{loss_pc_map} a). Djibouti is the worst-affected country in both cases with losses of up to $322.6$ kg per person in (\textit{India\_adap}, exceeding the loss in (\textit{India\_stat}. Many of the most affected countries tend to fare worse in terms of the availability of rice itself, but some countries can reduce the losses immensely. For example, the United Arab Emirates reduces losses from $58.8$ kg to $6.2$ kg per person, Kuwait decreases losses from $53.8$ kg to $49.7$ kg per person, and Oman not only eliminates its losses, but also manages to obtain a surplus of $11.3$ kg per person compared to the unshocked baseline. Conversely, countries that experience substantially worse rice availability include Djibouti, Benin (increasing from $98.7$ kg to $131.8$ kg per person), and Timor-Leste (increasing from $65.4$ kg to $155.6$ kg per person). Each of these countries triggered the adaptation mechanism, due to the high rice losses directly appearing at $t=1$ due to their high reliance on India for their rice imports. We can conclude that the compensatory efforts of countries do not necessarily lead to the mitigation of losses. Overall the rice losses are mitigated by about $1$ \%. The main effect of adaptation is only redistributory. Given the restriction on economic growth, the system must make do with what is left available, making it an entire mitigation of shocks impossible.\\

\begin{figure}[!h]
\centering
\includegraphics[width=0.9\linewidth]{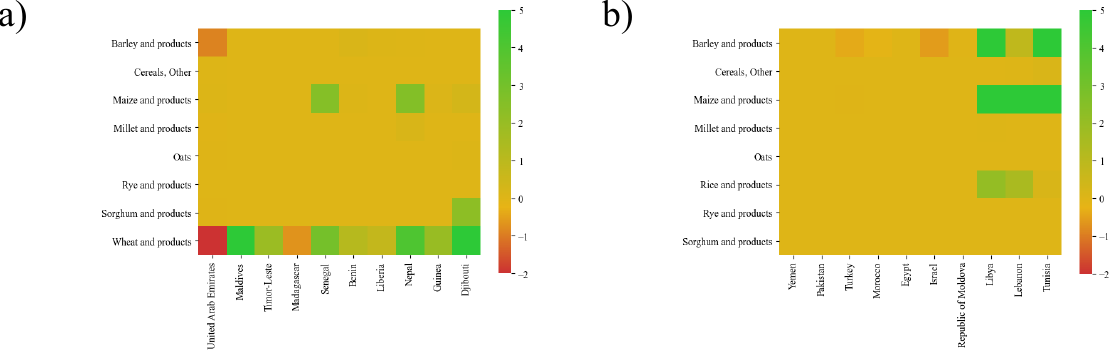}
\caption{Adaptation impact $(L_{a\rightarrow b}^{i\rightarrow j})_\text{stat}-(L_{a\rightarrow b}^{i\rightarrow j})_\text{adap}$ in kg per capita on possible substitute for top $10$ losers in the uncompensated scenario: a) Shock to Indian rice ($a=\text{\texttt{IND}}$, $i=\text{\texttt{Rice and products}}$,$j=\{Cereals\}$): The most popular substitute item is clearly wheat.  Only the United Arab Emirates do not see an increase in its availability with the adaptation.\\
b) Shock to Ukrainian wheat ($a=\text{\texttt{IND}}$, $i=\text{\texttt{Wheat and products}}$,$j=\{Cereals\}$): The three initially most affected countries — Tunisia, Lebanon, and Libya — import much more maize and barley.}
\label{substitutes}
\vspace{1cm}
\includegraphics[width=0.9\linewidth]{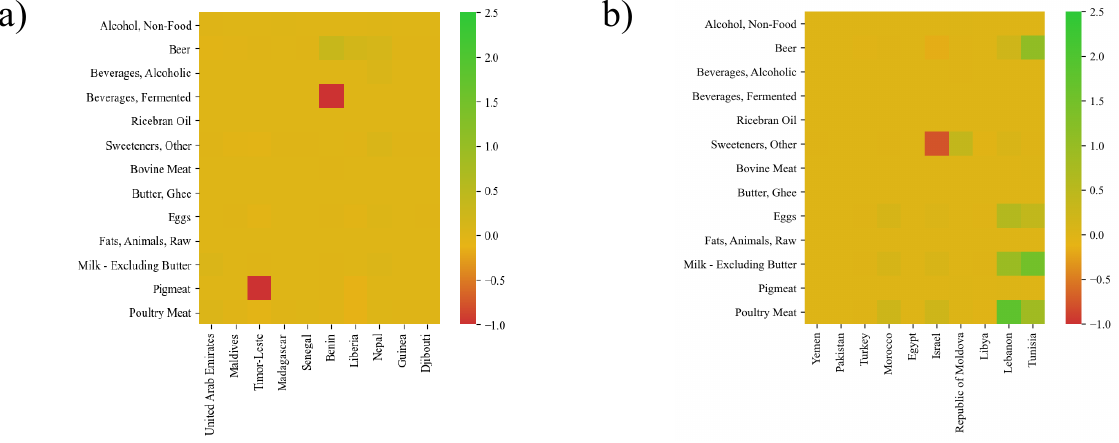}
\caption{Adaptation impact $(L_{a\rightarrow b}^{i\rightarrow j})_\text{stat}-(L_{a\rightarrow b}^{i\rightarrow j})_\text{adap}$ in kg per capita on possible secondary products for top $10$ losers in the uncompensated scenario: a) Shock to Indian rice ($a=\text{\texttt{IND}}$, $i=\text{\texttt{Rice and products}}$,$j=\{secondary\text{ } items\}$): Rice is not as important as an input product in production processes as wheat, and therefore the losses in available secondary products are rather small and do not change substantially with the adaptations, except for some singular losses in Liberia and Timor-Leste.\\
b) Shock to Ukrainian wheat ($a=\text{\texttt{UKR}}$, $i=\text{\texttt{Wheat and products}}$,$j=\{secondary\text{ } items\}$):: Especially Tunisia and Lebanon improve their availability of $j=$ \texttt{Poultry Meat}, \texttt{Milk - Excluding Butter}, \texttt{Eggs}, and \texttt{Beer}, indicating that the import of substitute items successfully support the efforts to mitigate losses.}
\label{secondary}
\end{figure}

The simulations also show that the rice shock can to a substantial extend be compensated by several countries by means of substitution. More concrete, positive correlations with all possible substitute cereals have been found for rice (see Figure \ref{trade_substitute_cereals_figure}), while wheat potentially provides the most impactful substitution, compare table \ref{trade_adaptation_substitutes_table} b). Appropriately, \ref{substitutes} a) shows increased wheat availability for almost all to most by the Indian rice shock affected countries. Interestingly, countries seem to employ deviating strategies: While the UAE aims to import more of the shocked product itself from its trade partners, others attempt to mitigate losses by importing substitute item.\\

When considering the impact of adaptations on secondary products, the shock to Indian rice shows little to no improvement as rice can in principle be used as an input product for the products mentioned in Figure \ref{secondary} a), but in reality rice is not as important as other cereals such as wheat or maize.\\
\textbf{Ukraine Wheat Scenario:}
In the simulation scenario, we shock $100$\% of Ukrainian wheat production ($a=\text{\texttt{UKR}}$, $i,j=\text{\texttt{Wheat and products}}$), resulting in Ukraine's wheat exports being reduced to nearly zero. The evaluation will focus solely on investigating losses in directly or indirectly affected countries other than Ukraine. Figure \ref{top50} b) illustrates the losses of wheat for the top $50$ most affected countries in \textit{Ukraine\_stat} for \textit{Ukraine\_adap} and \textit{Ukraine\_stat}, corresponding to Figure \ref{loss_pc_map} b). Tunisia emerges as the worst affected country in \textit{Ukraine\_stat} with losses of $122.5$ kg per capita, while Libya surpasses it in \textit{Ukraine\_adap} with losses of $187.5$ kg per capita. Most of the highly affected countries tend to fare better in terms of wheat availability, but a few countries' situations, including Libya's, deteriorate. Tunisia (increasing to $181.9$ kg per person) and Indonesia (rising from $11.8$ kg to $28.0$ kg per person) experience exacerbated losses as well. On the other hand, Lebanon reduces its losses from $108.2$ kg to $96.8$ kg per person, Greece from $18.1$ kg to $9.32$ kg per person, and Romania not only eliminates its losses, but also achieves a surplus of $1.4$ kg per person. Notably, the wheat shortages are generally smaller compared to the losses in rice in the other shock simulation, and the majority of countries perform better in \textit{Ukraine\_adap} compared to \textit{Ukraine\_stat}. The shock to Indian rice production is not only almost ten times larger in absolute size, but Indian rice also has about four times higher share of the global rice market ($32.6$\%) than Ukraine of the global wheat market ($9.3$\%). The Indian rice shock leads to between $82$ and $96$\% loss of rice supply for the top $10$ most affected countries in the uncompensated scenario, while the Ukrainian wheat shock only results in between $26$ and $72$\% loss of wheat supply for the top $10$ most affected countries in the uncompensated scenario.

Wheat is mostly substituted by barley and maize (figure \ref{substitutes} b)) which also agrees well with table \ref{trade_adaptation_substitutes_table} b).\\

For Ukrainian wheat the improvement for secondary products are more obvious than the ones in the 
Indian rice scenario. According to Table \ref{trade_adaptation_substitutes_table}, maize shows the highest correlation as a potential substitute for wheat. Figure \ref{secondary} b) reveals that the availability of maize increases, especially in Tunisia, Lebanon, and Libya. Barley, and to a lesser extent rice also seem to be popular choices for replacing wheat. Especially in Lebanon and Tunisia, but also Morocco and Egypt, we see substantial increases in the availability of animal products such as poultry meat, milk, eggs, and beer. As Libya and Tunisia experience higher wheat losses with compensation, they seemingly try to import alternative products, while other countries try managing the losses by importing wheat from other partners instead of increasing the import of other cereals. This highlights again the diverging strategies adopted by countries: While some target importing more of the shocked product itself, others attempt to mitigate losses by importing substitute products.\\

\newpage

\section{Commodity groups}\label{Commodity groups}
\begin{longtable}{ | l | l | } 
 \caption{Products and their respective product groups. Products within the same product group are considered as possible substitutes.}
 \label{table_product_group}\\
 \hline
 \textbf{Product group} & \textbf{Product} \\ 
 \hline
 \hline
 \multirow{4}{*}{Alcohol} & Beer\\
 & Beverages, Alcoholic\\
 & Beverages, Fermented\\
 & Wine\\
 \hline
 \multirow{1}{*}{Ethanol} & Alcohol, Non-food\\
 \hline
 \multirow{2}{*}{Fibre crops} & Cotton lint\\
  & Cottonseed\\
 \hline
 \multirow{9}{*}{Oil cakes} & Copra Cake\\
 & Cottonseed Cake\\
 & Groudnut Cake\\
 & Oilseed Cakes, other\\
 & Palmkernel, Cake\\
 & Rape and Mustard Cake\\
 & Sesameseed Cake\\
 & Soyabean Cake\\
 & Sunflowerseed Cake\\
 \hline
 \multirow{1}{*}{Oil crops} & Palm kernels\\
 \hline
 \multirow{3}{*}{Sugar, sweeteners} & Sugar (raw equivalent)\\
 & Sugar non-centrifugal\\
 & Sweeteners, other\\
 \hline
 \multirow{13}{*}{Vegetable oils} & Coconut Oil\\
 & Cottonseed Oil\\
 & Groundnut Oil\\
 & Maize Germ Oil\\
 & Oilcrops Oil, other\\
 & Olive Oil\\
 & Palm Oil\\
 & Palmkernel Oil\\
 & Rape and Mustard Oil\\
 & Ricebran Oil\\
 & Sesamesees Oil\\
 & Soyabean Oil\\
 & Sunflowerseed Oil\\
 \hline
 Fish & Fish, Seafood\\
 \hline
 \multirow{11}{*}{Live animals} & Asses\\
 & Buffaloes\\
 & Camelids, other\\
 & Cattle\\
 & Goats\\
 & Horses\\
 & Mules\\
 & Pigs\\
 & Poultry Birds\\
 & Rabbits and hares\\
 & Rodents, other\\
 \multirow{1}{*}{Live animals}& Sheep\\
 \hline
 \multirow{1}{*}{Animal fats} & Fats, Animal, Raw\\
 \hline
 \multirow{1}{*}{Eggs} & Eggs\\
 \hline
 \multirow{3}{*}{Hides, skins, wool} & Hides and skins\\
 & Silk\\
 & Wool (clean equivalent)\\
 \hline
 \multirow{1}{*}{Honey} & Honey\\
 \hline
 \multirow{6}{*}{Meat} & Bovine Meat\\
 & Meat, other\\
 & Mutton \& Goat Meat\\
 & Offals, edible\\
 & Pigmeat\\
 & Poulty Meat\\
 \hline
 \multirow{9}{*}{Cereals}& Barley and products \\ 
 & Cereals, other \\
 & Maize and products \\
 & Millet and products \\
 & Oats \\
 & Rice and products \\
 & Rye and products \\
 & Sorghum and products \\
 & Wheat and products \\
 \hline
 \multirow{3}{*}{Coffee, tea, cocoa} & Cocoa Beans and products \\ 
 & Coffee and products \\
 & Tea (including mate) \\
 \hline
 \multirow{6}{*}{Fibre crops}  & Abaca \\ 
 & Hard Fibres, other \\
 & Jute \\
 & Jute-Like Fibres \\
 & Sisal \\
 & Soft-Fibres, other \\
 \hline
 \multirow{1}{*}{Fodder crops} & Fodder crops \\
 \hline
 \multirow{1}{*}{Grazing} & Grazing \\
 \hline
 \multirow{4}{*}{Oil crops} & Coconuts (including Copra)\\
 & Groundnuts \\
 & Oil, palm fruit \\
 & Oilcrops, other \\
 \multirow{6}{*}{Oil crops} & Olives (including preserved) \\
 & Rape and Mustardseed \\
 & Seed cotton \\
 & Sesame seed \\
 & Soyabeans \\
 & Sunflower seed \\
 \hline
 \multirow{5}{*}{Roots and tubers} & Cassava and products\\
 & Potatoes and products\\
 & Roots, other\\
 & Sweet potatoes\\
 & Yams\\
 \hline
 \multirow{2}{*}{Sugar crops} & Sugar beet \\ 
 & Sugar cane \\ 
 \hline
 \multirow{2}{*}{Tabacco, Rubber} & Rubber \\ 
 & Tobacco \\
 \hline
 \multirow{23}{*}{Vegetables, fruit, nuts, pulses, spices} & Apples and products \\ 
 & Bananas \\
 & Beans \\
 & Citrus, other \\
 & Cloves \\
 & Dates \\
 & Fruits, other \\
 & Grapefruit and products \\
 & Grapes and products (excluding Wine) \\
 & Hops \\
 & Lemons, Limes and products \\
 & Nuts and products \\
 & Onions \\
 & Oranges, Mandarines \\
 & Peas \\
 & Pepper \\
 & Pimento \\
 & Pineapples and products \\
 & Plantains \\
 & Pulses, other and products \\
 & Spices, other \\
 & Tomatoes and products \\
 & Vegetables, other \\
 \hline
\end{longtable}
\newpage
\section{Human Development Index}\label{Human Development Index}
\begin{longtable}{ | l | l | l |} 
 \caption{Countries and their respective Human Development Index (HDI). Mind that not all of the countries and territories recognized by the UN are included in the model and are therefor missing in the table.}
 \label{table_human_development_index}\\
 \hline
 \textbf{HDI category} & \textbf{Country} & HDI  \\ 
 \hline
 \hline
 \multirow{44}{*}{Very high development} & Switzerland & 0.962\\
 & Norway & 0.961\\
 & Iceland & 0.959\\
 & China, Hong Kong SAR & 0.952\\
 & Australia & 0.951\\
 & Denmark & 0.948\\
 & Sweden & 0.947\\
 & Ireland & 0.945\\
 & Germany & 0.942\\
 & Netherlands & 0.941\\
 & Finland & 0.940\\
 & Singapore & 0.939\\
 & Belgium & 0.937\\
 & New Zealand & 0.937\\
 & Canada & 0.936\\
 & Luxembourg & 0.930\\
 & United Kingdom & 0.929\\
 & Japan & 0.925\\
 & Republic of Korea & 0.925\\
 & United States of America & 0.921\\
 & Israel & 0.919\\
 & Malta & 0.918\\
 & Slovenia & 0.918\\
 & Austria & 0.916\\
 & United Arab Emirates & 0.911\\
 & Spain & 0.905\\
 & France & 0.903\\
 & Cyprus & 0.896\\
 & Italy & 0.895\\
 & Estonia & 0.890\\
 & Czech Republic & 0.889\\
 & Greece & 0.887\\
 & Poland & 0.876\\
 & Bahrain & 0.875\\
 & Lithuania& 0.875\\
 & Saudi Arabia& 0.875\\
 & Portugal& 0.866\\
 & Latvia & 0.863\\
 & Croatia & 0.858\\
 & Chile & 0.855\\
 & Qatar & 0.855\\
 & Slovakia & 0.848\\
 & Hungary & 0.846\\
 & Argentina & 0.842\\
 \multirow{19}{*}{Very high development} & Turkey & 0.838\\
 & Montenegro & 0.832\\
 & Kuwait & 0.831\\
 & Brunei Darussalam & 0.829\\
 & Russian Federation & 0.822\\
 & Romania & 0.821\\
 & Oman & 0.816\\
 & Bahamas & 0.812\\
 & Kazakhstan & 0.811\\
 & Trinidad and Tobago & 0.810\\
 & Costa Rica & 0.809\\
 & Uruguay & 0.809\\
 & Belarus & 0.808\\
 & Panama & 0.805\\
 & Malaysia & 0.803\\
 & Georgia & 0.802\\
 & Mauritius  & 0.802\\
 & Serbia & 0.802\\
 & Thailand & 0.800\\
 \hline
 \multirow{33}{*}{High human development} & Albania & 0.796\\
 & Bulgaria & 0.795\\
 & Grenada & 0.795\\
 & Barbados & 0.790\\
 & Antigua and Barbuda & 0.788\\
 & Sri Lanka & 0.785\\
 & Bosnia and Herzegovina & 0.782\\
 & Saint Kitts and Nevis & 0.780\\
 & Iran (Islamic Republic of) & 0.777\\
 & Ukraine & 0.774\\
 & North Macedonia & 0.773\\
 & China, mainland & 0.770\\
 & Dominican Republic & 0.768\\
 & Republic of Moldova & 0.767\\
 & Cuba & 0.764\\
 & Peru & 0.762\\
 & Armenia & 0.759\\
 & Mexico & 0.758\\
 & Brazil & 0.754\\
 & Colombia  & 0.752\\
 & Saint Vincent and the Grenadines & 0.751\\
 & Maldives & 0.747\\
 & Algeria & 0.745\\
 & Azerbaijan & 0.745\\
 & Turkmenistan & 0.745\\
 & Ecuador & 0.740\\
 & Mongolia  & 0.739\\
 & Egypt & 0.731\\
 & Tunisia & 0.731\\
 & Fiji & 0.730\\
 & Suriname & 0.730\\
 & Uzbekistan & 0.727\\
 & Dominica & 0.720\\
 \multirow{12}{*}{High human development} & Jordan & 0.720\\
 & Libya & 0.718\\
 & Paraguay & 0.717\\
 & Saint Lucia & 0.715\\
 & Guyana & 0.714\\
 & South Africa & 0.713\\
 & Jamaica & 0.709\\
 & Samoa & 0.707\\
 & Gabon & 0.706\\
 & Lebanon & 0.706\\
 & Indonesia & 0.705\\
 & Viet Nam & 0.703\\
 \hline
 \multirow{37}{*}{Medium human development} & Philippines & 0.699\\
 & Botswana & 0.663\\
 & Bolivia (Plurinational State of) & 0.692\\
 & Kyrgyzstan & 0.692\\
 & Venezuela (Bolivarian Republic of) & 0.691\\
 & Iraq & 0.686\\
 & Tajikistan & 0.685\\
 & Belize & 0.683\\
 & Morocco & 0.683\\
 & El Salvador & 0.675\\
 & Nicaragua & 0.667\\
 & Cabo Verde & 0.662\\
 & Bangladesh & 0.661\\
 & India & 0.633\\
 & Ghana & 0.632\\
 & Guatemala & 0.627\\
 & Kiribati & 0.624\\
 & Honduras & 0.621\\
 & Sao Tome and Principe & 0.618\\
 & Namibia & 0.615\\
 & Lao People's Democratic Republic & 0.607\\
 & Timor-Leste & 0.607\\
 & Vanuatu & 0.607\\
 & Nepal & 0.602\\
 & Eswatini & 0.597\\
 & Cambodia & 0.593\\
 & Zimbabwe & 0.593\\
 & Angola & 0.586\\
 & Myanmar & 0.585\\
 & Syrian Arab Republic & 0.577\\
 & Cameroon & 0.576\\
 & Kenya & 0.575\\
 & Congo & 0.571\\
 & Zambia & 0.565\\
 & Solomon Islands & 0.564\\
 & Papua New Guinea & 0.558\\
 & Mauritania & 0.556\\
 & Côte d'Ivoire & 0.550\\
 \hline
 \multirow{2}{*}{Low human development}& United Republic of Tanzania & 0.549\\
 & Pakistan & 0.544\\
 \multirow{30}{*}{High human development} & Togo & 0.539\\
 & Haiti & 0.535\\
 & Nigeria & 0.535\\
 & Rwanda & 0.534\\
 & Benin & 0.525\\
 & Uganda & 0.525\\
 & Lesotho & 0.514\\
 & Malawi & 0.512\\
 & Senegal & 0.511\\
 & Djibouti & 0.509\\
 & Sudan & 0.508\\
 & Madagascar & 0.501\\
 & Gambia & 0.500\\
 & Ethiopia & 0.498\\
 & Eritrea & 0.492\\
 & Guinea-Bissau & 0.483\\
 & Liberia & 0.481\\
 & Democratic Republic of the Congo & 0.479\\
 & Afghanistan & 0.478\\
 & Sierra Leone & 0.477\\
 & Guinea & 0.465\\
 & Yemen & 0.455\\
 & Burkina Faso & 0.449\\
 & Mozambique & 0.446\\
 & Mali & 0.428\\
 & Burundi & 0.426\\
 & Central African Republic & 0.404\\
 & Niger & 0.400\\
 & Chad & 0.394\\
 & South Sudan & 0.385\\
\hline
\multirow{14}{*}{not categorized} & Belgium-Luxembourg & --\\
& China, Macao SAR & --\\
& China, Taiwan Province of & --\\
& Czechoslovakia & --\\
& Democratic People's Republic of Korea & --\\
& French Polynesia & --\\
& Netherlands Antilles & --\\
& New Caledonia & --\\
& Puerto Rico & --\\
& RoW & --\\
& Serbia and Montenegro & --\\
& Somalia & --\\
& USSR & --\\
& Yugoslav SFR & --\\
\hline
\end{longtable}
\newpage

\end{document}